\documentclass[aps,pra,superscriptaddress,twocolumn,reprint,amsmath,amssymb,graphicx]{revtex4-2}

\usepackage[utf8]{inputenc}
\usepackage[T1]{fontenc}

\newif\ifstructure
\structuretrue

\newcommand{\beq}{\begin{equation}}
\newcommand{\eeq}{\end{equation}}
\newcommand{\bea}{\begin{eqnarray}}
\newcommand{\eea}{\end{eqnarray}}
\providecommand{\abs}[1]{\left\lvert#1\right\rvert}

\providecommand{\bra}[1]{\langle #1 \rvert}
\providecommand{\ket}[1]{\lvert #1 \rangle}

\newcommand{\ketbra}[2]{\left| {#1} \right\rangle\left\langle {#2}\right|}

\newcommand{\wirp}{\omega_{\text{ir,probe}}}
\newcommand{\wirr}{\omega_{\text{ir,ref}}}
\newcommand{\xuv}{\text{xuv}}
\newcommand{\ir}{\text{ir}}
\newcommand{\irp}{\text{ir,probe}}
\newcommand{\irr}{\text{ir,ref}}
\newcommand{\fano}{\text{fano}}

\newcommand{\erf}{\text{erf}}

\usepackage{afterpage}
\usepackage{empheq}

\usepackage{soul}

\usepackage{subcaption}
\usepackage{tikz}
\usepgflibrary{arrows}
\usetikzlibrary{decorations}
\usetikzlibrary{decorations.pathmorphing,patterns}
\usetikzlibrary{scopes}
\usetikzlibrary{shapes, matrix}
\usepackage{xcolor}

\usepackage{appendix}

\begin{document}

\title{A multidimensional approach to quantum state tomography of photoelectron wavepackets}

\author{H. \surname{Laurell}}
\affiliation{Material Sciences Division, Lawrence Berkeley National Laboratory, Berkeley, California 94720, USA}
\affiliation{Department of Chemistry, University of California, Berkeley, California, 94720, USA}
\affiliation{Department of Physics, Lund University, Box 118, 22100 Lund, Sweden}
\author{J. \surname{Baños Gutiérrez}}
\affiliation{Instituto de Quimica, Universidad Nacional Autonoma de Mexico, Circuito Exterior, Ciudad Universitaria, Alcaldía Coyoacán C.P. 04510, Ciudad de Mexico}
\author{A. \surname{L'Huillier}}
\affiliation{Department of Physics, Lund University, Box 118, 22100 Lund, Sweden}
\author{D. \surname{Busto}}
\affiliation{Department of Physics, Lund University, Box 118, 22100 Lund, Sweden}
\affiliation{Physikalisches Institut, Albert-Ludwigs-Universität Freiburg, Hermann-Herder-Straße 3, 79104 Freiburg, Germany}
\author{D. \surname{Finkelstein-Shapiro}}
\email{daniel.finkelstein@iquimica.unam.mx}
\affiliation{Instituto de Quimica, Universidad Nacional Autonoma de Mexico, Circuito Exterior, Ciudad Universitaria, Alcaldía Coyoacán C.P. 04510, Ciudad de Mexico}

\begin{abstract}
There is a growing interest in reconstructing the density matrix of photoelectron wavepackets, in particular in complex systems where decoherence can be introduced either by a partial measurement of the system or through coupling with a stochastic environment.
To this end, several methods to reconstruct the density matrix, quantum state tomography protocols, have been developed and tested on photoelectrons ejected from noble gases following absorption of extreme ultraviolet (XUV) photons from attosecond pulses. 
It remains a challenge to obtain model-free, single scan protocols that can reconstruct the density matrix with high fidelities. Current methods require extensive measurements or involve complex fitting of the signal. Efficient single-scan reconstructions would be of great help to increase the number of systems that can be studied.
We propose a new and more efficient protocol that is able to reconstruct the continuous variable density matrix of a photoelectron in a single time delay scan. 
It is based on measuring the coherences of a photoelectron created by absorption of an XUV pulse using a broadband infrared (IR) probe that is scanned in time and a narrowband IR reference that is temporally fixed to the XUV pulse. We illustrate its performance for a Fano resonance in He as well as mixed states in Ar arising from spin-orbit splitting. We show that the protocol results in excellent fidelities and near-perfect estimation of the purity. 
\end{abstract}

\maketitle

\section{Introduction}

The discovery of high-order harmonic generation (HHG) \cite{McPherson1987,Ferray1988} and the synthesis of attosecond light pulses has enabled the study of electron dynamics in real time. 
The dynamics of ionization have been measured \cite{Isinger2017,Busto2022,Schultze2010,Klunder2011}, through interferometric measurements of a photoelectron following absorption of extreme ultraviolet (XUV) and infrared (IR) pulses. 
Both the spectral phase variation and the amplitude of the photoelectron can be determined. For a fully coherent state, the reconstruction of the photoelectron wavefunction is possible up to a global phase.
The photoelectron wavepacket structure depends on the manifold of states available. Above the ionization threshold this manifold consists of discrete autoionizing states and continuous free electron levels with different angular momenta. 
This rich structure leads to many types of non-trivial ionization phenomena, for example Fano or shape resonances  \cite{Fano1961,Kaldun2016,Finkelstein2017,Hammerland2024,Ahmadi2022,Zhong2020,Nandi2020}.
\newline

Several factors either from the experiment or intrinsic to the system may introduce decoherence. As a result, the wavefunction no longer describes the system accurately and we must use the density matrix formalism instead. 
Experimental imperfections degrade the measured coherences, as illustrated in experiments in neon \cite{Bourassin2020}. 
Additionally, incomplete measurements on an entangled system can also result in decoherence. This has been illustrated in angle-integrated measurements of autoionizing wave packets in helium where radial and angular degrees of freedom can be entangled \cite{Busto2022}. Similarly, in the presence of ion-photoelectron entanglement, averaging over the ionic degree of freedom also introduced decoherence \cite{Laurell2022,Laurell2023,Vrakking2021,Koll2022}. 
%
In the above cases, a complete characterization of the photoelectron quantum state requires methods to reconstruct the density matrix $\rho$. \newline

Different successful reconstructions of the density matrix have been put forward.
Bourassin-Bouchet et al. used the mixed-FROG scheme to reconstruct the density matrix of photoelectrons ejected from neon. The measurement consists of using 
an attosecond pulse train to generate photoelectron wavepackets with components at different energies. Then, the absorption of a time delayed high-intensity IR laser pulse results in interferences due to different multi-photon transitions leading to the same final state.  
An iterative retrieval algorithm is then used to reconstruct the photoelectorn density matrix.
A purity of 0.11 was retrieved in the particular case of neon and assigned to a decoherence introduced by the spectrometer response, and fluctuations in the XUV-IR delay \cite{Bourassin2020}. 

Laurell et al. developed a protocol using a delayed bichromatic IR probe, called KRAKEN. The idea is that two monochromatic IR pulses, characterized by a frequency difference $\delta \omega$, select two energy levels of the electron wavepacket and interfere them in a final state, thus probing their coherence. 
The experiment is run multiple times with different frequency combinations of the bichromatic pulse to probe different pairs of states \cite{Laurell2022}. The principle of using a very selective probing pulse simplifies the interpretation and the transformation from the signal to density matrix population or coherence. 
Recently, this protocol has been demonstrated experimentally in He and Ar \cite{Laurell2023}. 
Seven spectrograms with different values of the shear $\delta \omega$ were sufficient to obtain a reconstruction with excellent agreement with theory. However, this protocol is very time-consuming and requires interpolation methods to fill in the signal between different shears. \newline

Ideally, a quantum state tomography protocol is robust to experimental noise, can be performed efficiently in a short time and requires minimal processing to obtain the density matrix. If a protocol is to measure coherences between arbitrary pairs of levels, then it needs a label for each level, which we are distinguishing by its energy. In standard photoelectron spectroscopy, there is one energy dimension associated to the detection of the photoelectron energy, so that the quantum state tomography protocol needs to create an additional energy dimension, usually through judiciously chosen time delays between pulses. 
Multidimensional spectroscopy using XUV and NIR pulses has been successfully implemented to probe bound states in Ar \cite{Cao2016,Marroux2018}. Similarly, excitation resolved spectroscopy with phase-locked XUV twin pulses has been applied to bound states in He as well as long-lived resonances in Ar \cite{Wituschek2020}. Optical two-dimensional spectroscopy or Raman spetroscopy analogues have been proposed in the XUV to probe core electron dynamics \cite{Mukamel2013,Rahav2010}. However, these sequences are not suitable for interrogating continuum states where transitions to the final detected state only occur  during XUV-IR pulse overlap. 

In this work, we propose a new version of the KRAKEN protocol that uses the full spectrum of a broadband IR pulse to reconstruct the density matrix in a single time delay scan. 
We refer to this variation of the KRAKEN protocol as rainbow KRAKEN, in analogy to rainbow RABBIT technique \cite{Gruson2016,Busto2018}, which is an energy resolved version of the standard RABBIT (reconstruction of attosecond beating by interference of two-photon transitions \cite{Paul2001}).
As in the original protocol, we record electron spectra as a function of the delay between XUV and IR pulses, but replace the bichromatic probe by a combination of a broadband probe and a narrowband reference, both in the IR spectral range. We begin by motivating the physical basis of the protocol and describe how coherences are encoded in Fourier space for different pulse sequences. We derive analytical formulas for the measured signal and use them to describe the data processing needed to go from the measurement to the density matrix. After analyzing the sources of error and the differences with the original KRAKEN experiment, we illustrate our method in two model cases, a photoelectron created in the vicinity of the $2s2p$ resonance in He and one created in the unstructured continuum of Ar with two different ionic states, 3p$^5$ $^2$P$_{3/2}$ and 3p$^5$ $^2$P$_{1/2}$, where ion-electron entanglement introduces decoherence when only the electron is measured, resulting in a mixed density matrix. \newline

\section{Theoretical description of the pulse sequence}

\subsection{Encoding an indirect energy dimension in a time delay}

A density matrix contains the populations of the different quantum states, which are labelled by their quantum numbers, and the coherences between the states.
For a photoelectron, the quantum numbers are the kinetic energy, the angular momentum, the magnetic quantum number and the electron spin. Tracing over the angular, magnetic and spin degrees of freedom results in a reduced density matrix solely described by the kinetic energy of the photoelectron, which is what we consider in the following. 

\begin{figure*}[htb!]
    \centering
    \includegraphics[width=0.9\textwidth]{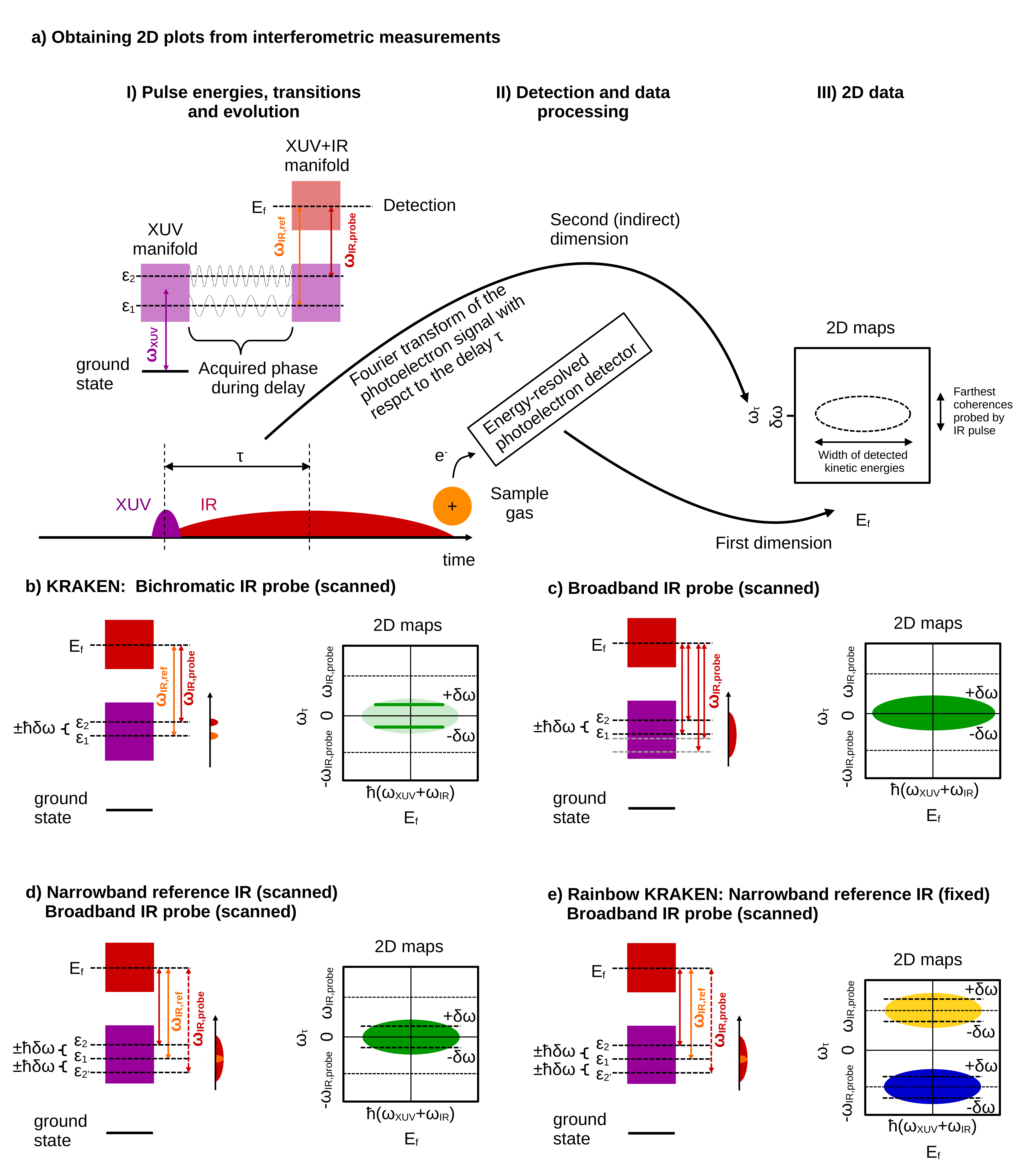}
    \captionsetup{singlelinecheck=off, format=plain, width=\textwidth} 
    \caption{a) Structure of a general pulse sequence using one XUV and one (possibly structured) IR pulse. The first XUV pulse creates a photoelectron in the XUV manifold. At time $\tau$, an IR pulse promotes the photoelectron into a higher manifold, the XUV+IR manifold, where it is detected. 
    The delay causes an accumulated phase for each energy state $\ket{\varepsilon}$ which can be used to label it. 
    The final photoelectron kinetic energy is taken as the first dimension (x-axis) while the Fourier transform of the signal with respect to the delay time $\tau$ is taken as the second dimension (y-axis). The time and spectral properties of the IR pulse, whose options we explore in panels b-e,  will influence the selectivity and multiplexing ability of the experiment. For clarity, we have not depicted the final photoelectrons that have interacted with a single component of the IR pulse as these can be filtered out, as described in the main text.
    b) The KRAKEN experiment uses a delayed bichromatic IR probe that unambiguously selects levels at a fixed energy difference and interferes them in a final state. 
    This occupies two narrow regions in Fourier space at $\delta \omega = \pm (\wirp-\wirr)$.  
    c) A delayed IR broadband pulse can probe coherences of levels separated by a continuum of energy differences in a single time scan but mixes them in the final signal indiscriminately. 
    d) A broadband IR probe and narrowband IR reference both with a delay $\tau$ can select contributions to levels equidistant in energy to a reference level. 
    e) Rainbow KRAKEN: A broadband IR probe scanned in time $\tau$ and a narrowband IR reference pulse fixed in time can unambiguously distinguish energy levels.}
    \label{fig:pulse_sequences_encoding}
\end{figure*}

To determine coherences between pairs of states, a two-dimensional measurement is required. Standard photoelectron experiments measure the kinetic energy of electrons, providing only one-dimensional data. This type of measurement does not allow us to infer the coherences between pairs of states at different energies within the reduced density matrix. To achieve this, we need a second dimension that also measures the photoelectron energy. This is accomplished using interferometric techniques.
A prototypical two-photon experiment is shown in Figure \ref{fig:pulse_sequences_encoding}.a. A XUV pulse, with central frequency $\omega_\xuv$ and width $\sigma_\xuv$ prepares a wavepacket in at time $\tau=0$. The possible energies are labelled by $\varepsilon_i$. 
An IR pulse delayed by a delay $\tau$, with center frequency $\omega_\ir$ and spectral width $\sigma_\ir$ promotes the photoelectron to a higher final XUV+IR energy labelled by $E_f$ where it is detected. 
Interferometric techniques with time-delayed pulses naturally have two dimensions: the first one is directly provided by the detection method with energy resolution (taken as the x-axis), and the other is the delay $\tau$ which can be converted into an energy $\hbar \omega_\tau$ through a Fourier transform (taken as the y-axis).  
We can follow a similar analysis as in multidimensional NMR, IR and visible light spectroscopy \cite{Mukamel1995,Jonas2003,Hamm1998} to obtain an intuitive picture of the effect of the pulse sequence.
Due to the time delay, each state accumulates a different phase. The difference in accumulated phase is related to their energy difference, so that the signal coming from the resulting interference in a higher lying state will be modulated by this energy difference (Fig. \ref{fig:pulse_sequences_encoding}.a).
We thus obtain a 2D map that correlates different states (Fig. \ref{fig:pulse_sequences_encoding}.a).
In attosecond photoionization experiments, however, the two-photon signal only exists during the overlap of the XUV and IR pulses \cite{JimenezGalan2016} so that no phase accumulation due to free evolution can occur. 
In this case, the free evolution is replaced by the delay $\tau$ between the XUV and IR fields. 

The interferometric scheme should allow the measurement of all coherences between all occupied levels of the continuum that form the wavepacket. 
There are a number of ways in which the encoding of the second dimension can be achieved depending on the energy and temporal properties of the pulses. 

Figure \ref{fig:pulse_sequences_encoding} presents four possible pulse sequences. To understand the effect of these sequences, we focus on two states of a photoelectron in the XUV continuum with energies $\varepsilon_1$ and $\varepsilon_2$. 
A complete wavefunction is composed of a continuum of levels but since coherence is a pairwise property, it is sufficient to consider two levels.
We consider only a pure state, but the treatment provided here can be easily generalized to mixed states.  
We form a wavefunction from a linear combination of these two levels $\ket{\Psi(\tau)} = c(\varepsilon_1)e^{-i \varepsilon_1 \tau/\hbar} \ket{\varepsilon_1} + c(\varepsilon_2)e^{-i \varepsilon_2 \tau/\hbar} \ket{\varepsilon_2}$, with $\abs{c(\varepsilon_1)}^2+\abs{c(\varepsilon_2)}^2=1$.  
We begin by analyzing the KRAKEN protocol and continue by investigating possible sequences for single scan protocols. \newline

\textit{KRAKEN: bichromatic IR pulse}. The KRAKEN protocol uses a bichromatic IR pulse with frequencies $\wirr$, $\wirp$. The final continuum state $\ket{E_f}$ can be reached by two interfering quantum paths with intermediate states $\ket{\varepsilon_1}$ and $\ket{\varepsilon_2}$, obeying the energy conservation condition $E_f = \varepsilon_1 + \hbar \wirr = \varepsilon_2 + \hbar \wirp$ (Figure \ref{fig:pulse_sequences_encoding}.b). 
Although absorption of the XUV pulse creates a photoelectron spanning a large range of energies, only two energies $(\varepsilon_1,\varepsilon_2)$  are promoted to the final state $\ket{E_f}$. 
Assuming that the initial photoelectron is created at time $\tau = 0$, and that the two spectral components of the IR pulse arrive at a time $\tau$ after the XUV pump, the final measured signal will be 
\begin{equation}
\begin{split}
S_K(E_f,\tau) & \propto | \Psi(\tau)|^2 \\ 
& \propto |c(\varepsilon_1)|^2 + |c(\varepsilon_2)|^2 \\ 
& + c(\varepsilon_1)c^*(\varepsilon_2)e^{-i (\varepsilon_1-\varepsilon_2) \tau/\hbar} \\
& + c^*(\varepsilon_1)c(\varepsilon_2)e^{i (\varepsilon_1-\varepsilon_2) \tau/\hbar}
\end{split}
\label{eq:two-level-expression}
\end{equation}
where the proportionality sign is valid as long as the continuum-continuum transitions can be considered independent from the energy of the intermediate state. 
The signal will have constant and oscillating terms as function of the delay $\tau$. This can be visualized more clearly with the Fourier transform of the signal $\mathcal{F}_{\tau}\{ S_K(E_f,\tau) \}(\omega_\tau)$, where for a function $f(\tau)$, 
$\mathcal{F}_{\tau} \{ f(\tau) \}(\omega_\tau) \equiv \tilde{f}(\omega_\tau) = \int d\tau f(\tau) e^{-i\omega_\tau \tau} $ where $\omega_{\tau}$ is the conjugate frequency to the delay time $\tau$. The signal occupies three regions: the constant terms appear at $\omega_{\tau}=0$ (with complex amplitudes $\abs{c(\varepsilon_1)}^2$ + $\abs{c(\varepsilon_2)}^2$) while the beating terms will appear at the difference frequencies $\omega_{\tau} = \pm \delta \omega = \pm (\varepsilon_2-\varepsilon_1)/\hbar$ (with intensities $c(\varepsilon_1)c^*(\varepsilon_2) = \rho_{\xuv}(\varepsilon_1,\varepsilon_2)$ and $c^*(\varepsilon_1)c(\varepsilon_2)= \rho_{\xuv}(\varepsilon_2,\varepsilon_1)$). These latter two components are the coherences of the density matrix. 

In the case of ionization by a broad XUV pulse, a continuum of states is populated. As a result, a continuum of final states is reached after interaction with the bichromatic IR pulse.
Each detected final kinetic energy $E_f$ has associated components $\omega_\tau$, resulting in a complex valued surface with the x-axis labelling the photoelectron kinetic energy $E_f$ and the y-axis the conjugate frequency $\omega_\tau$. 
In the standard KRAKEN approach, most of the two dimensional Fourier map is empty - only the spectral information at $\omega_\tau=\pm\delta\omega$ is used to reconstruct a subdiagonal of the density matrix. Multiple spectrograms obtained for different spectral shears $\delta\omega$ are necessary to sample different subdiagonals of the density matrix. 
An efficient protocol will encode as much information as possible in the Fourier map. In the following we discuss alternative schemes that make a more efficient use of the Fourier space to encode more information in a single spectrogram. \newline

\textit{XUV pulse, broadband IR pulse.} To reconstruct the density matrix in a single delay scan, we need to simultaneously encode the coherences between all levels of the photoelectron wavepacket. This can be achieved by including all bichromatic pairs at once in the form of a broadband IR probe (Figure \ref{fig:pulse_sequences_encoding}.c). When we carry out the Fourier transform of the signal with a single delayed broadband IR pulse, a large Fourier space area becomes accessible. However, for a given final level with kinetic energy $E_f$, and a frequency component $\omega_\tau$, there is now a continuum of pairs of levels that contribute, introducing an ambiguity that precludes reconstruction of the density matrix. The broad bandwidth of the IR pulse is capable of inducing interferences between many pairs of levels $\varepsilon_i$ but does not provide discrimination. \newline

\clearpage

\textit{Rainbow-KRAKEN: interfering broadband and narrowband IR pulses.} We overcome the lack of discrimination coupled to the use of a broadband pulse by the re-introduction of a narrowband reference IR pulse at frequency $\wirr$, and filter the signal to select only the  interference that comes from photoelectrons that interacted with the narrowband reference and the broadband probe (Figure \ref{fig:pulse_sequences_encoding}.d). 
For a final detected photoelectron kinetic energy $E_f$ and a given Fourier frequency $\omega_\tau$, the contributing pairs are set by the reference at $\varepsilon_1 = E_f-\hbar \wirr$, within the broad bandwidth of the IR probe. Since the different levels will acquire a different phase after the delay $\tau$, a given frequency $\omega_{\tau}$ can only have contributions from levels such that $\varepsilon_2-\varepsilon_1 = \hbar \omega_{\tau}$, or $\varepsilon_1-\varepsilon_{2'} = \hbar \omega_{\tau}$. This is because each pair of levels will appear in Fourier frequency at positive and negative frequencies of equal magnitude. This discrimination is advantageous compared to the case without an IR reference, although it is still not enough to reconstruct the density matrix because the correlations between the states $\varepsilon_1$ and $\varepsilon_2$, and those between $\varepsilon_{2'}$ and $\varepsilon_{1}$ are mixed in the same final signal. We can also see this from the expansion of the simple two-level system wavefunction $\Psi$ (Eq. \eqref{eq:two-level-expression}): a pair of energy levels will have contributions at positive and negative frequencies $\delta \omega_\tau$ so that levels equidistant from the reference $\varepsilon_1 = (E_f-\hbar \wirr)$ are indistinguishable.

We separate these two contributions by fixing the IR reference pulse in time with respect to the XUV pulse and only scan the delay of the IR probe pulse. This shifts all frequencies in the Fourier space at $\pm \wirp$ and lifts the remaining ambiguity.  
%
We can understand the shift in Fourier space as follows: the signal acquires a delay-dependent phase related to the energy of the level of the XUV manifold. 
The interference terms between states $c(\varepsilon_1)c^*(\varepsilon_2)$ or $c^*(\varepsilon_1)c(\varepsilon_2)$ involve complex conjugates in a way that the phases of the two interfering states are subtracted.
When, in the axes corresponding to Fig. \ref{fig:pulse_sequences_encoding}.b-d, we measure an interference between levels $\ket{\varepsilon_1}$ and $\ket{\varepsilon_2}$, the total modulation of the signal is done at a frequency corresponding to the energy difference between these levels. These frequencies are on the order of $\pm(\wirr-\wirp)$.
When the delay of the reference IR pulse is locked to the XUV, the contribution to the signal from the state that has interacted with the reference does not have a delay-dependent phase. 
The delay-dependent phase of the signal coming from interfering states is $(E_f-
\varepsilon_2)\tau /\hbar$ and is on the order of $(\pm \wirp \pm \sigma_\irp)\tau$ because of the energy conservation argument $E_f \approx \hbar \omega_\xuv + \hbar \wirp$. This leads to a 2D map such as in Fig. \ref{fig:all_Fourier_encodings}.e with two components centered at $\pm \wirp$.

Let us consider a signal due to a component of the broadband IR probe, with frequency $\wirp$, and the reference IR pulse, with frequency $\wirr$.
The components such that $\wirp-\wirr>0$, will appear at positive frequencies in the Fourier conjugate $\omega_\tau$, while those such that $\wirp-\wirr<0$ will appear at negative frequencies, where we have used $\wirp$ to indicate one frequency component of the broadband probe and not its center frequency. 

\begin{figure*}[htb!]
    \centering
    \includegraphics[width=1.0\textwidth]{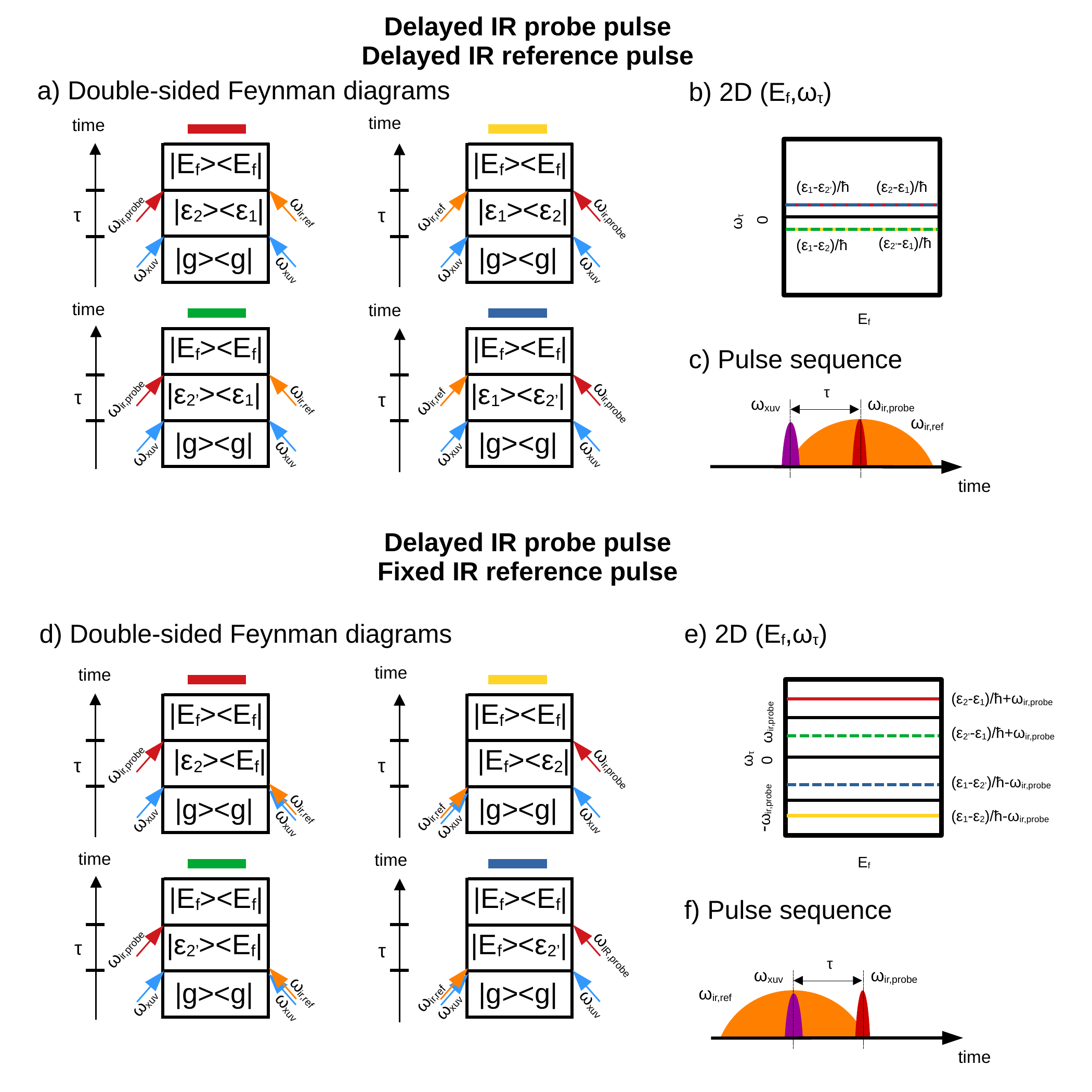}
    \caption{a) Feynman diagrams for the rainbow-KRAKEN experiment with delayed IR reference pulse and delayed IR probe pulse, b) position of each diagram's contribution in the 2D ($E_f,\omega_\tau$) map and c) pulse sequence. d, e and f show the same diagrams as a,b,c, respectively, for the experiment with a fixed IR reference pulse. The Feynman diagrams as written are valid for the conditions of an impulsive XUV and IR probe pulse and a frequency selective IR reference pulse.}
    \label{fig:Feynman_diagrams}
\end{figure*}

We can better visualize this with the double-sided Feynman diagrams representation for the sequence (Figure \ref{fig:Feynman_diagrams}, \cite{Mukamel1995}). These diagrams keep track of the state of the density matrix after each light-matter interaction. They start with a ground state density matrix $\ketbra{g}{g}$ and interactions are added as incoming (absorption) or outgoing (emission) arrows, with forward time evolution read from bottom to top of the diagram. They can be very useful to understand the encoding in Fourier space. For a delayed probe and delayed reference IR pulses (Fig. \ref{fig:Feynman_diagrams}.a,b,c), during the time delay $\tau$ the density matrix is in state $\rho = \ketbra{\varepsilon_i}{\varepsilon_j}$ with $i,j=1,2,2'$. There is a phase factor associated with this density matrix of $e^{-i(\varepsilon_i-\varepsilon_j)\tau/\hbar}$, whose argument is on the order of $-i\sigma_{\irp} \tau$. 
After Fourier transforming, we obtain a signal at $\omega_{\tau} = -(\varepsilon_i-\varepsilon_j)/\hbar$, very close of and around $\omega_\tau = 0$. There are four contributions appearing in only two distinct frequencies. When we fix the reference pulse and allow for a delayed IR probe (Fig. \ref{fig:Feynman_diagrams}.d,e,f), the density matrix evolves during the time delay in the state $\rho=\ketbra{\varepsilon_i}{E_f}$ or $\rho=\ketbra{E_f}{\varepsilon_i}$ for $i=2,2'$, with an accumulated phase $e^{\pm i(\varepsilon_i-E_f)\tau/\hbar}$, whose argument is on the order of $\pm i \wirp \tau$. The signals appear at $\omega_\tau = \pm (\varepsilon_i-E_f)$. The four diagrams are encoded in distinct frequencies in Fourier space. We note that we have followed a reasoning appropriate for bound states, however an analogous phase due to the electric field of the IR probe pulse is obtained for the states of the continuum that require immediate excitation into the XUV+IR manifold after they have been excited to the XUV manifold.
%
%

The rainbow-KRAKEN makes use of the entire Fourier space accessible with a given IR probe bandwidth and unambiguously encodes the correlations needed to reconstruct the density matrix in a single measurement. \newline

\subsection{Analytical expressions for the rainbow-KRAKEN interferogram.} 

We present in this section the expressions relate  a measured signal to a density matrix. We start with a heuristic derivation that captures the essential physics of the protocol, followed by an exact calculation based on the two-photon transition probability amplitude of absorbing an XUV and an IR photon. In both derivations we will strictly focus on the contributions that arise from the interference between probability amplitudes of absorption of an IR reference pulse and an IR probe pulse. It is understood that the detector will also measure a signal of photoelectrons that have only interacted with the reference or the probe alone. However, this component can be subtracted. It also appears in a different region of the Fourier space so that it can be simply filtered out. \newline

\textbf{Heuristic derivation of the Rainbow-Kraken signal.} Let us consider a photoelectron that is created by absorption of an XUV photon at time $\tau=0$. Its state can be described by a one-photon density matrix $\rho_{\xuv}$. We apply the rainbow-KRAKEN IR pulses to obtain a two-photon density matrix $\rho_{\xuv+\ir}$. The observed signal for a final kinetic energy $E_f$ is:

\begin{equation}
    \tilde{S}_{\text{RK}}(E_f,\hbar \omega_{\tau}) = \mathcal{F}_\tau \left( \bra{E_f} \rho_{\xuv + \ir}(\tau) \ket{E_f} \right)(\omega_\tau)
\end{equation}

In the limit of a monochromatic reference IR pulse and an impulsive IR probe, we can relate the signal to the initial density matrix by (see Appendix A for calculations)
\begin{equation}
\begin{split}
    \tilde{S}_{RK}(E_f,\hbar \omega_\tau) &= \tilde{S}^{(+)}_{RK}(E_f, \hbar \omega_\tau) \\
    &+ \tilde{S}^{(-)}_{RK}(E_f,\hbar \omega_\tau) \\ 
    \tilde{S}^{(+)}_{RK}(E_f,\hbar \omega_\tau) & \approx  
    A_{\text{probe}} A_{\text{ref}} |\mu_{c}|^2 \\
    &\times \rho_\xuv( E_f - \hbar\omega_\tau ,E_f - \hbar \wirr) \\
    \tilde{S}^{(-)}_{RK}(E_f,\hbar \omega_\tau) & \approx  
    A_{\text{probe}} A_{\text{ref}} |\mu_{c}|^2 \\
    & \times \rho_\xuv^*(E_f + \hbar\omega_\tau, E_f - \hbar \wirr)
\end{split}
\label{eq:heuristic}
\end{equation}
where $A_\text{probe}$ and $A_\text{ref}$ are the electric field amplitudes of the IR probe and reference pulse respectively, $\mu_c$ is the energy independent transition dipole moment from going to the intermediate continuum states $\{ \ket{\varepsilon} \}$ to the final detected continuum states $\{ \ket{E} \}$. 
The signal is composed of two terms, $\tilde{S}^{(-)}_{RK}$ and $\tilde{S}^{(+)}_{RK}$, which are the negative $\omega_\tau$ and positive $\omega_\tau$ regions explained in the previous section (Figure \ref{fig:pulse_sequences_encoding}.e). The density matrix $\rho_\xuv$ is composed of states with kinetic energies obeying the approximate relation $\varepsilon \approx E_f - \hbar \wirp$. Consequently, from equation \eqref{eq:heuristic} we see that the signals will appear for the condition $E_f - \hbar \wirp \approx E_f \pm \hbar \omega_\tau$, that is at $\omega_{\tau} \approx \pm \wirp$. 
The density matrices are parameterized by the argument $\varepsilon_2^{\pm} = E_f \mp \hbar\omega_\tau$ and $\varepsilon_1 = E_f - \hbar \wirr$. We can continuously measure $E_f$ and $\omega_\tau$, so that we then have access to the desired reconstruction $\rho_{\xuv}(\varepsilon_2,\varepsilon_1)$. The signals at positive and negative $\omega_\tau$ frequencies have the same information and the same signal to noise ratio so that we can consider only one of these. \newline 

In addition to the assumption of a monochromatic reference and a broad bandwidth probe, we have also assumed that the final signal (Eq.~\eqref{eq:heuristic}) can be decomposed into two separated, sequential one-photon absorption processes. This is not the case, and we must look instead at the two-photon transition amplitude to obtain the exact signal. As we will see, this will result in small deviations from the one-photon density matrix $\rho_\xuv$, but the full calculation will allow us to correct for finite-pulse effects exactly. \newline

\textbf{Explicit exact expressions from the two-photon absorption cross-section.} The Rainbow-KRAKEN 2D $(E_f,\hbar \omega_\tau)$ maps can be constructed from the interference between the two-photon transition amplitude using an XUV pulse and a fixed narrowband IR pulse $\mathcal{A}_{\wirr}(E_f, \tau=0)$ and that of an XUV pulse and a time-delayed broadband IR pulse, $\mathcal{A}_{\wirp}(E_f, \tau)$,
\begin{equation}
\begin{split}
    \tilde{S}_{\text{RK}}(E_f,\omega_{\tau}) & = \mathcal{F}_{\tau} \left \{ \abs{ \mathcal{A}_{\wirp}(E_f,\tau) + \mathcal{A}_{\wirr}(E_f,0)}^2 \right. \\ 
    & \left. - \abs{\mathcal{A}_{\wirp}(E_f,\tau)}^2-\abs{\mathcal{A}_{\wirr}(E_f,0)}^2 \right \}(\omega_{\tau}) \\
    &= \tilde{\mathcal{A}}_{\wirp}(E_f,\omega_\tau) \mathcal{A}^*_{\wirr}(E_f,0) \\
    & + \tilde{\mathcal{A}}^*_{\wirp}(E_f,-\omega_\tau) \mathcal{A}_{\wirr}(E_f,0)
\end{split}
\label{eq:total_signal_time}
\end{equation}
We have used the identity $\mathcal{F}_{\tau} \{ f^*(\tau) \}(\omega_\tau) = \tilde{f}^*(-\omega_\tau)$. We can identify $\tilde{S}_{\text{RK}}^{(+)}(E_f,\omega_{\tau}) = \tilde{\mathcal{A}}_{\wirp}(E_f,\omega_\tau) \mathcal{A}^*_{\wirr}(E_f,0)$ and $\tilde{S}_{\text{RK}}^{(-)}(E_f,\omega_{\tau}) = \tilde{\mathcal{A}}^*_{\wirp}(E_f,-\omega_\tau) \mathcal{A}_{\wirr}(E_f,0)$. \newline



If we assume that all pulses have Gaussian envelopes, it is possible to obtain analytical forms for both 
$\tilde{\mathcal{A}}_{\wirp}(\omega_\tau)$ and $\mathcal{A}_{\wirr}(0)$ so as to derive the expression for the signal and the steps needed to reconstruct the density matrix precisely \cite{JimenezGalan2016}.
\begin{figure}
    \centering
    \includegraphics[width=0.5\textwidth]{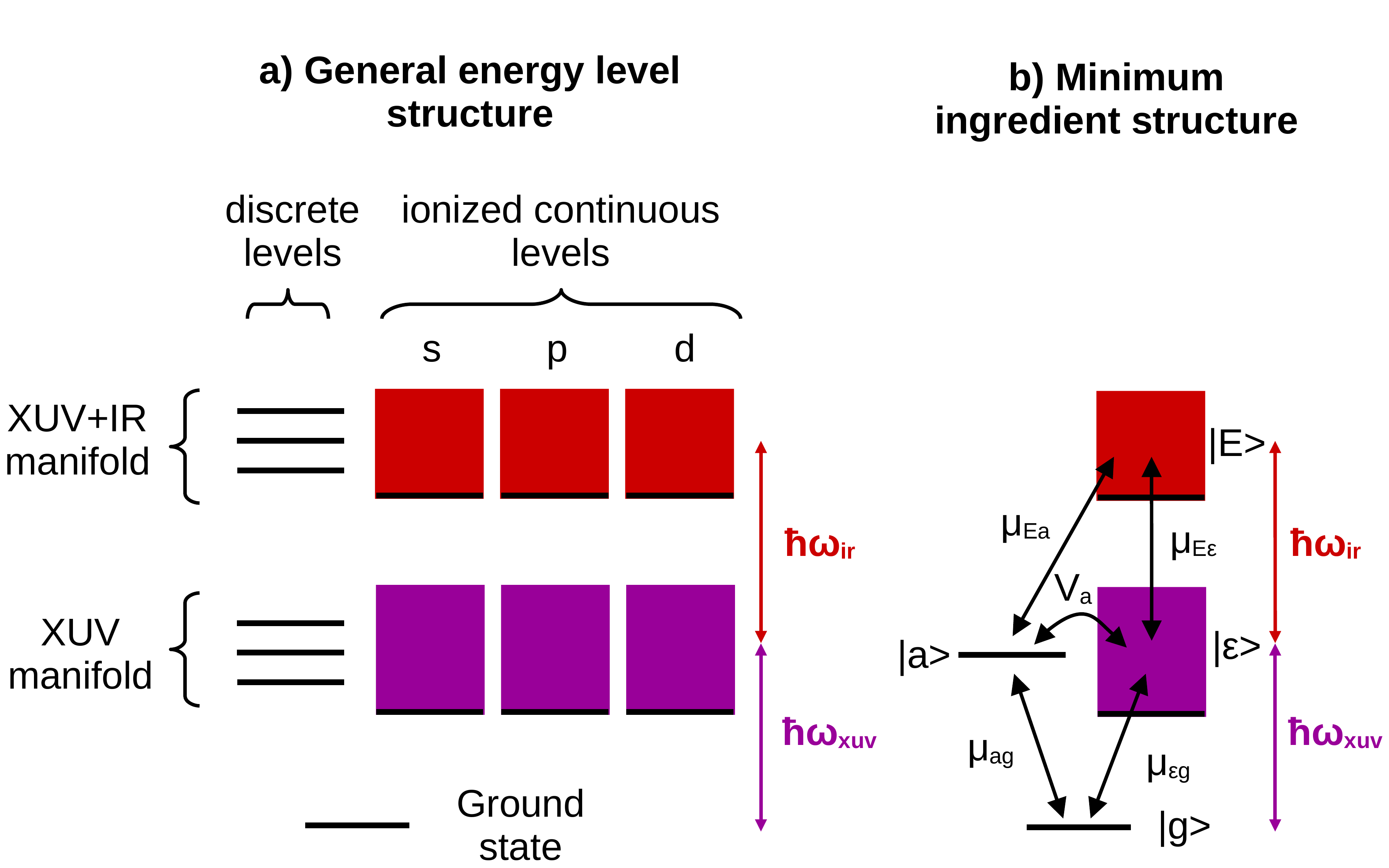}
    \caption{a) General energy level structure of the XUV and XUV+IR manifolds consisting of a combination of discrete resonances and several continua with different angular momentum. b) A representative minimum ingredient model consisting of one discrete and one continuum for the XUV manifold, and one flat continuum for the XUV+IR manifold.}
    \label{fig:general_energy_levels}
\end{figure}
Exact analytical expressions require an explicit structure of the energy levels accessible after absorption of an XUV photon and the energy levels accessible after further absorption of an IR photon. Fig. \ref{fig:general_energy_levels}.a shows a general energy structure for both manifolds consisting of combinations of discrete and continuum states. We start by deriving the equations in the case of He, from which it can be easily extended to other atoms. Radiative transitions couple the ground state to the XUV manifold, and the XUV manifold to the XUV+IR manifold. Selection rules for electrons initially in an $s$ shell dictate that they are excited to the continuous states of the $p$ shell. 
Subsequently, the IR photon further excites them to the continuum states of the $s$ and $d$ shells, where they are detected. The discrete levels are coupled to the continuum states via non-radiative transitions. 
We can simplify our derivation by focusing on a minimum ingredient model (the general case is solved in Appendix C). Far from the ionization threshold, it is usually possible to tune the experiment so that the XUV+IR manifold is approximately flat, i.e. so that the discrete levels are far way from the detected states $\{ \ket{E_f} \}$. We can also simplify the number of continua in our model. In the XUV manifold we only need to consider states with angular momentum $p$ because they are the only ones accessible radiatively from the ground state. We include coupled discrete levels, however we only need to consider one since qualitatively, adding more discrete levels does not change the complexity of the expressions. If we consider a flat XUV+IR manifold, i.e. no discrete levels in the final state, then the final photoelectron radial and angular degrees of freedom are not entangled, and so it is enough to have a single effective continuous manifold \cite{Busto2022}. We then arrive at the energy structure of Figure  \ref{fig:general_energy_levels}.b). The coupling between a discrete level and a continuum has been described by Fano \cite{Fano1961}. It has a destructive interference for a specific energy in the spectrum. This distinctive feature is very useful to illustrate the different steps in going from a recorded interferogram to the density matrix. In this Fano structure we consider that the discrete level lies at an energy $\hbar \omega_{ag}$ and is connected to the continuum states $\ket{\varepsilon}$ through a configuration interaction $V_a$. We label the radiative transition dipole moment from state $j$ to $i$, $\mu_{ij}$.
To describe the effect of the coupling between the discrete state and the continuum, Fano introduced the factor
\begin{equation}
f_\fano(\epsilon, q) = \frac{\epsilon+q}{\epsilon+i}   
\label{eq:standrd_fano}
\end{equation}
where $\epsilon = (\varepsilon-\hbar \omega_{ag})/\Gamma_a $ is the detuning, $\Gamma_a = \pi V_a^2$ the width of the resonance and where $q  \equiv q_{ag} = \frac{\mu_{ag}}{\pi V_a \mu_{\varepsilon g}}$. 
The corresponding density matrix is \cite{Laurell2022}
\begin{equation}
 \begin{split}
    \rho_{\xuv}(\epsilon_2,\epsilon_1) &= A  G(\varepsilon_2/\hbar - \omega_\xuv , \sigma_\xuv) \\
    & \times G(\varepsilon_1/\hbar - \omega_\xuv ,\sigma_\xuv) \\
    & \times f_\fano(\epsilon_2, q_{ag}) f_\fano^*(\epsilon_1, q_{ag}) \\
    \end{split}
    \label{eq:theoretical_rho}
\end{equation}
where $A$ is a normalization constant, $G(x,\sigma) = e^{-\frac{x^2}{2\sigma^2}}$ is a Gaussian envelope with a normalized maximum amplitude. This is the density matrix whose reconstruction we will illustrate. 
The exact expression of the Fourier transform of the Rainbow-KRAKEN interferogram signal is 
decomposed into two components (see Appendix B for details of the derivation)
\begin{equation}
\begin{split}
    \tilde{S}_{RK}^{(+)}(E_f,\omega_\tau) & =
    I_0 G(\delta_{\text{ref}}, \sigma_\xuv) M_{\text{probe}}^{(+)}(E_f,\omega_\tau) \\
    & \times f^*_\fano(\epsilon_{E_f}, q_{\text{ref}}) f_\fano(-\epsilon_\tau^{(+)}, q_{\text{probe}}), \\
    \tilde{S}_{RK}^{(-)}(E_f,\omega_\tau) & = I_0 G(\delta_{\text{ref}}, \sigma_\xuv) M_{\text{probe}}^{(-)}(E_f,\omega_\tau) \\
    & \times f_\fano(\epsilon_{E_f}, q_{\text{ref}}) f^*_\fano(-\epsilon_\tau^{(-)}, q_{\text{probe}})
\end{split}
\label{eq:exact_analytical_expression}
\end{equation}
where $I_0$ is a constant, $\delta_i = \omega_\xuv + \omega_\text{IR,i} - E_f/\hbar$, for $i=$ ref, probe, and $q_\text{ref}$, $q_\text{probe}$ are modified asymmetry parameters (Appendix B). We use the detunings $\epsilon_{E_f} = \frac{E_f-\hbar \wirr - \hbar \omega_{ag}}{\Gamma_a}$ and $\epsilon_{\tau}^{(\pm)} = \frac{\pm \hbar \omega_\tau - (E_f - \hbar \omega_{ag} )}{\Gamma_a}$. The explicit form of $M^{(\pm)}_{\text{probe}}$ is,
\begin{equation}
\begin{split}
 & M^{(\pm)}_{\text{probe}}(E_f,\omega_\tau)  = \\
& \exp \left(-\frac{\sigma_t^2}{2}(\pm \omega_\tau -\omega_\irp + \frac{\sigma_\irp^2}{\sigma^2}\delta_{\text{probe}})^2 \right) \\
&\times \exp \left( -\frac{\delta_{\text{probe}}^2}{2\sigma^2} \right) \times \frac{1}{\pm\omega_\tau} 
\end{split}
\label{eq:Mprobe}
\end{equation}
where $\sigma=\sqrt{\sigma_\xuv^2 +\sigma_\irp^2}$ and $\sigma_t=\sqrt{\sigma_\xuv^{-2} +\sigma_\irp^{-2}}$,
 determines the region in Fourier space where the signal appears. $\tilde{S}_{RK}^{(+)}(E_f,\omega_\tau)$ appears at positive frequencies and $\tilde{S}_{RK}^{(-)}(E_f,\omega_\tau)$ for negative frequencies. 


We can see in the structure of Eq.~\eqref{eq:exact_analytical_expression} that we recover the expected Fano profiles albeit with slightly different asymmetry parameters as well as forms of the detuning. These are dependent on $E_f$ and $\omega_\tau$ and are determined from our labelling scheme. The profiles are modulated by functions that reflect the IR spectral envelope and depend on the sequence of pulses applied. 
We can use the relabelling of energies $\varepsilon_2^{\pm} = E_f \mp \hbar\omega_\tau$ and $\varepsilon_1 = E_f - \hbar \wirr$ to 
bring the expression closer to a density matrix with two independent energy axes.
Structurally, the expression is analogous to the theoretical density matrix we want to reconstruct (Eq. \eqref{eq:exact_analytical_expression}), or to the heuristic derivation of the Rainbow-KRAKEN (Eq. \eqref{eq:heuristic}) in that we have a modulation of the signal reflecting the pulses and an energy dependent function depending on the energy structure of the intermediate manifold, but the form of the functions is not exactly the same. We need to understand better the prefactor, modulation and profiles in order to propose a transformation between them. \newline

\textit{Modulation functions of the pulse sequence.} The signal is modulated by two functions. $G(\delta_\text{ref},\sigma_\xuv)$ represents up to a constant the imprint of the XUV spectrum on the initial photoelectron wavepacket, and is part of the theoretical expression for the density matrix Eq. \eqref{eq:theoretical_rho}: we cannot excite an infinitely broad wavepacket since $\sigma_\xuv$ is finite. $M^{(\pm)}_{\text{probe}}(E_f,\omega_\tau)$ is a more complicated weighting function that depends both on the width of the XUV pulse as well as that of the IR probe. It reflects the possible coherences that can be measured with a given IR probe. 
The largest distance in energy whose coherence we need to probe is set by the energy width of the excited wavepacket, which is limited by the XUV spectrum. However, the maximum distance between levels that we can probe can be smaller if the IR probe pulse bandwidth is not broad enough. 
Also, the signal scales with  the strengths of the electric field of the IR probe pulse and IR reference pulse. Photoelectrons that have interacted with frequency components towards the tail of the Gaussian pulse will have a weaker signal than those that have interacted with the center of the distribution. Because of this, the intensity of the signal does not automatically reflect the density matrix we wish to reconstruct.
In the limit of an infinitely broad IR probe - which is an ideal probe - this extra modulation is lost and we can verify that we recover the modulation imposed by the XUV spectrum alone, that is $\lim_{\sigma_\text{IR,probe} \to \infty} M^{(\pm)}_{\text{probe}}(E_f,\omega_\tau) = \frac{1}{\pm\omega_\tau}\exp \left(-\frac{(\pm \omega_{\tau}+ \omega_\xuv-E_f/\hbar)}{2\sigma_\xuv^2} \right) \equiv \pm\frac{1}{\omega_\tau} G(\delta_{\omega_\tau}^{(\pm)},\sigma_\xuv)$ with $\delta_{\omega_{\tau}}^{(\pm)} =  \pm \omega_{\tau}+ \omega_\xuv-E_f/\hbar$. Since we want a probe-independent measurement, any dependence on the probe and reference pulses on the final signal has to be removed. The modulation functions presented are a result of the protocol and valid for arbitrarily complex structures of the levels after absorption of an XUV photon as long as the final detected continuum $\ket{E_f}$ is flat and the continuum-continuum transition dipole moments are independent of the energy of the states.\newline

\textit{Energy-dependent transition probability amplitudes.} The profile has also the contribution that we are after, an energy dependence that is intrinsic to the particular structure of the energy levels. In the case of a Fano structure, we can reexpress it using the Fano form as long as we use the following complex asymmetry parameters
\begin{equation}
    \begin{split}
        q_{\text{ref}} & = q_{ag}(1 + \text{Re}(\Delta_\text{ref})/q_{ag}) + i\text{Im}(\Delta_\text{ref}) \\
        q_{\text{probe}} & = q_{ag}(1 + \text{Re}(\Delta_\text{probe})/q_{ag}) + i\text{Im}(\Delta_\text{probe})
    \end{split}
\end{equation}
where $\Delta_\text{ref}$ and $\Delta_{\text{probe}}$ are small complex-valued functions (see Eqs. \eqref{eq:Delta_ref} and \eqref{eq:Delta_probe}). The asymmetric parameters are modified because $q_{ag}$ describes the interference processes arising from a one-photon transition while the Rainbow-KRAKEN is a two-photon transition. 




\begin{figure*}[htb!]
    \centering
    \includegraphics[width=0.9\textwidth]{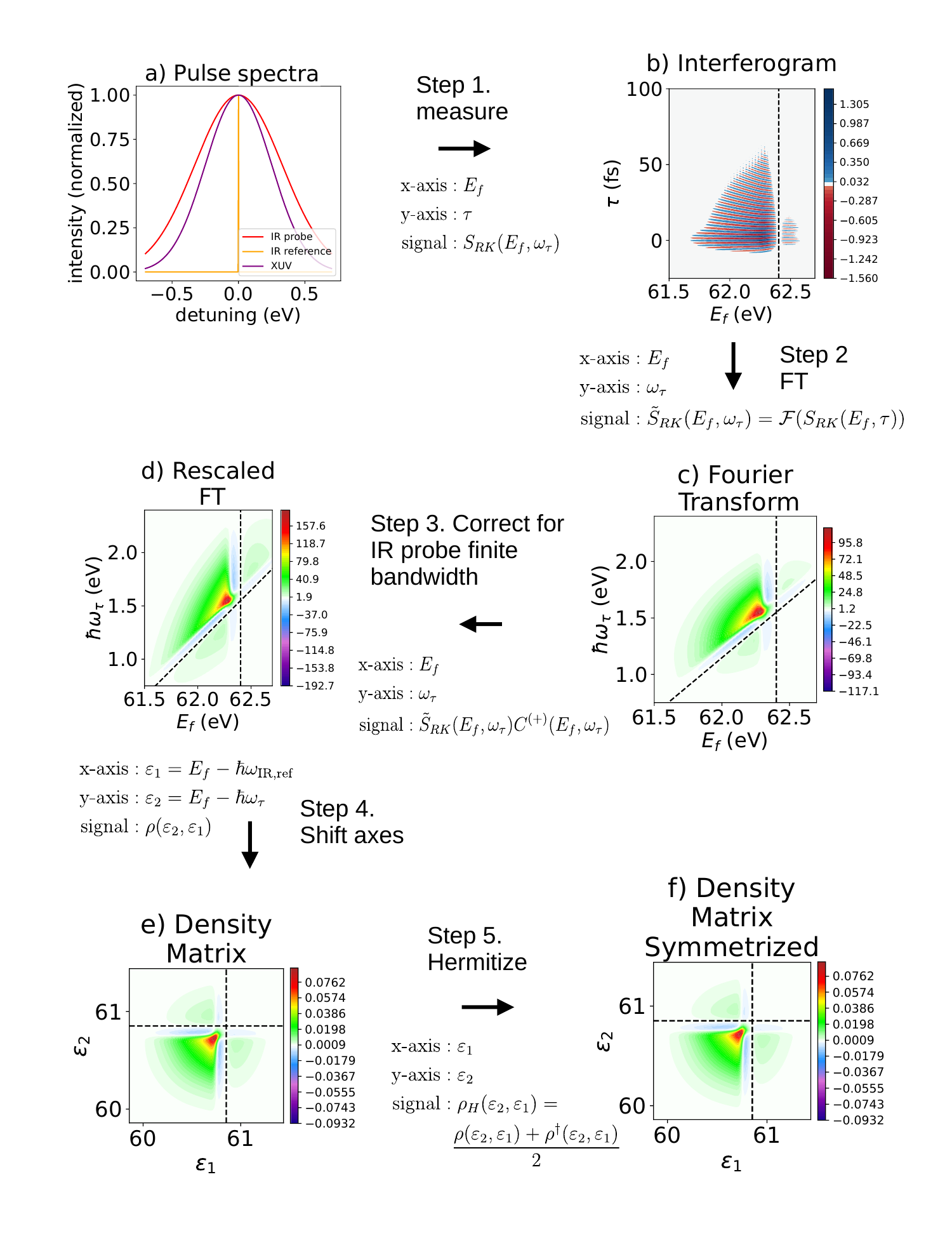}
    \captionsetup{singlelinecheck=off, format=plain, width=\textwidth} 
    \caption{Procedure for transforming the time trace into the density matrix. a) Pulse envelopes for the XUV, IR reference and IR probe pulses. The x-axis is the detuning with respect to the center frequency. b) Time trace for rainbow-KRAKEN (interference terms only). The vertical dotted line is the destructive interference at $E_f = \hbar (\omega_{ag} + \wirr)-\Gamma_a q_{ag}$.  c) Real part of the Fourier transform of the interferogram for the 2s2p transition of helium. In addition to the vertical line marking the destructive interference, there is a tilted line that occurs at $\hbar \omega_{\tau} = \hbar \omega_{ag}-E_f -\Gamma_a q_{ag} $. d) Same real part of the Fourier transform as in c) corrected for the finite bandwidth of the probe IR pulse using Eq. \eqref{eq:correction}. e) Real part of the density matrix obtained by re-scaling the x- and y-axis so as to depict the XUV manifold. f) Final density matrix after using the Hermitization procedure $\rho_H = \frac{1}{2}(\rho+\rho^{\dagger})$.}
    \label{fig:procedure}
\end{figure*}

\clearpage

\section{Reconstruction of the density matrix.} 

%
We describe the transformations to be carried out to transform the Fourier map described by Eq. \eqref{eq:exact_analytical_expression} into a density matrix. This transformation is composed of shifting energy axes, removing the dependence on the probe and reference pulses, and rescaling of the axes. We illustrate the procedure in Figure \ref{fig:procedure} for the case of the 2s2p Fano resonance in helium which features an asymmetry factor of $q=-2.77$. The destructive interference serves as a good spectral feature to more clearly keep track of what each transformation to the signal is doing. \newline

\textit{Step 1. Measure an interferogram (Fig. \ref{fig:procedure}.a,b).} We begin with a simulated interferogram with XUV and IR pulses shown in Figure \ref{fig:procedure}.a appropriately shifted to compare their widths. Figure \ref{fig:procedure}.b shows the interferogram, with only contributions from the interference of reference and probe pulses (Eq. \eqref{eq:total_signal_time}). The x-axis is the kinetic energy $E_f$ of the detected photoelectron while the y-axis is the time delay $\tau$ between the XUV pulse and the broadband IR probe pulse. 
We can first observe that the photoelectron is centered at an energy $E_f \approx \hbar(\omega_\xuv + \omega_\irr) - I_p$, where we have set the ionization potential $I_p=0$. Fano profiles have a destructive interference minimum for $\epsilon+q = 0$, and the interferogram shows this clear zero intensity region where expected, at $E_f = \hbar \omega_{ag} -\Gamma_a/q_{ag} + \hbar \wirr$ (Figure \ref{fig:procedure}.b, dotted black line). Along the y-axis we can see the broad feature of direct ionization, present during pulse overlap, followed by the narrower decay of the resonance at later times. \newline

\textit{Step 2. Fourier transform.} Fourier transforming the interferogram yields the 2D $(E_f,\hbar \omega_{\tau})$ map, with signals appearing in the conjugate frequency positions  $\omega_{\tau} = \pm (\wirp -\frac{\sigma_\irp^2}{\sigma^2} \delta_\text{probe})$ which we establish from the maximum of the exponential term of the modulation function $M^{(\pm)}$ (Figure \ref{fig:procedure}.c). \newline

\begin{figure}
    \centering
    \includegraphics[width=0.5\textwidth]{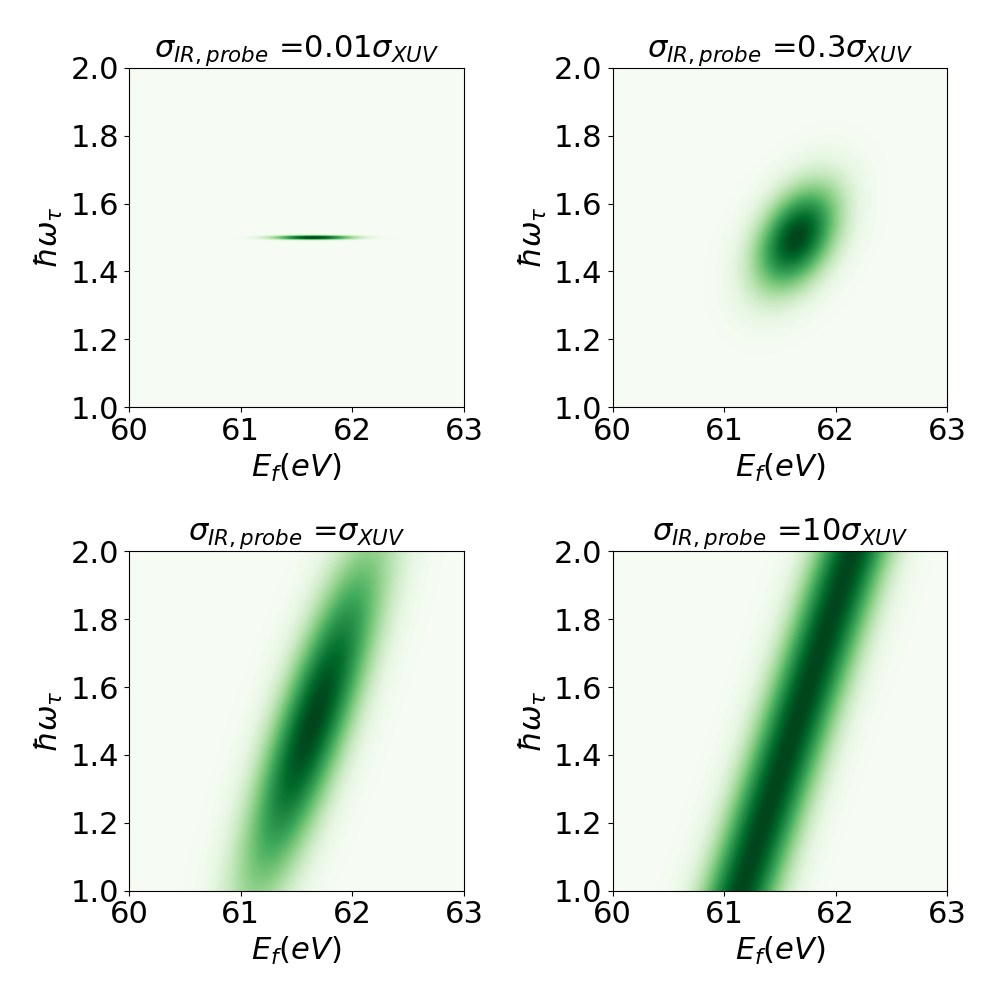}
    \caption{Envelope function $M^{(\pm)}_{\text{probe}}(E_f, \omega_{\tau})$ for different values of the probe bandwidth $\sigma_\irp$ marks the accessible Fourier frequencies. For very large bandwidths the span of accessible frequencies starts to be limited by the energy spread of the wavepacket excited by the XUV pulse and not the function $M^{(\pm)}_{\text{probe}}$. We have chosen $\hbar \wirp = 1.55$ eV.}
    \label{fig:M_function}
\end{figure}

\textit{Step 3. Correct the modulation of the IR probe spectrum.} 
We have to remove the intensity modulation imposed by the IR probe spectrum, and restore the natural envelope arising from the XUV spectrum. 
For this, we multiply the signal by 

\begin{equation}
\begin{split}
C^{\pm}(E_f,\omega_\tau) =& \left[ M^{(\pm)}_{\text{probe}}(E_f,\omega_{\tau})+\zeta \right]^{-1} \\
& \times \omega_\tau \lim_{\sigma_{\text{IR,probe}} \to \infty} M^{(\pm)}_{\text{probe}}(E_f,\omega_{\tau}) \\
=& \left[ M^{(\pm)}_{\text{probe}}(E_f,\omega_{\tau})+\zeta \right]^{-1} G( \delta_{\omega_\tau}^{(\pm)}, \sigma_\xuv)
\end{split}
\label{eq:correction}
\end{equation}
We have introduced the small number $\zeta$ to avoid numerical divergences from dividing by very small numbers. In practice we have chosen $\zeta = 0.001$. We show in Figure \ref{fig:procedure}.c such a rescaling where the Fourier transformed signal looks broader than in Figure \ref{fig:procedure}.b. There is naturally a limit after which this procedure cannot work. We discuss these limitations below. 
Figure \ref{fig:M_function} shows the modulation function $M^{(\pm)}_\text{probe}$ from Eq. \eqref{eq:Mprobe} for different values of the IR probe bandwidth. The span along $\hbar \omega_\tau$ for a given detection energy $E_f$ represents the distance in energy between levels whose coherence can be probed. For very narrow spectra, there are almost no coherences probed, and as we approach the limit of an infinitely spectrally broad pulse the distance is not limited by the IR pulse any longer but by the levels that can be populated by the XUV pulse.  \newline

\textit{Step 4. Shift and rescale the x and y-axis.} It is evident from Figure \ref{fig:procedure}.d that one destructive interference feature is vertical while the other has a tilt (see Eq. \eqref{eq:theoretical_rho}). The tilt can be easily read from the expression of the effective detuning $\epsilon_\tau$. We define two new axis which will correspond to the energies of levels in the XUV manifold, $\varepsilon_1 = E_f-\hbar\wirr$ and $\varepsilon_2 = E_f-\omega_\tau$ (Figure \ref{fig:procedure}.d). After this correction the destructive interference appears as a strictly horizontal feature in the rescaled map, and the energy associated to $\hbar \omega_\tau$ now directly corresponds to the manifold $\{\ket{\varepsilon} \}$ accessible after absorption of an XUV photon. The shift of the x-axis by $\hbar \wirr$ labels the same manifold. 
It is instructive to trace the origin of the tilt of the destructive interference in the original Fourier transform. It relates to the encoding of the energy along the indirect dimension. For this let us consider a final energy $E_{f,1}$, and a position of destructive interference that we define at $\varepsilon_D = \hbar \omega_D$. The destructive interference state is encoded at a frequency $\omega_{\tau,1} = E_{f,1}/\hbar - \omega_D$. For another photoelectron energy such that $E_{f,2}>E_{f,1}$, the encoding will now be at a different higher frequency $\omega_{\tau,2} = E_{f,2}/\hbar - \omega_D > \omega_{\tau,1}$, even though it refers to the same level. So, a given state is encoded at larger and larger Fourier conjugate frequencies as we go higher and higher in detected photoelectron energy. It is a consequence of how we encode the second dimension in the frequency $\omega_\tau$ and explains the tilt. \newline

\textit{Step 5. Hermitize the density matrix.} The last step is enough to obtain the density matrix. However, the errors incurred in assuming that a two-photon transition amplitude gives the one-photon density matrix are not the same for both axes (i.e., the functions $\Delta_\text{probe}$ and $\Delta_\text{ref}$ are not the same). As a consequence, the reconstructed density matrix is not necessarily Hermitian. 
As a final step we enforce Hermiticity by defining the Hermitian density matrix $\rho_H \equiv \frac{1}{2}(\rho+\rho^{\dagger})$. Since the signals at positive and negative frequencies $\tilde{S}_{RK}^{(\pm)}$ have the same information, we keep only the positive frequency term. The operation to obtain $\rho_H$ removes any numerical errors accumulated during the previous steps, and averages the functions $\Delta_\text{ref}$ and $\Delta_\text{probe}$. The final expression then becomes
\begin{equation}
\begin{split}
    \rho_H(\varepsilon_2,\varepsilon_1) &= I_0 G(\varepsilon_2/\hbar-\omega_\xuv,\sigma_\xuv) G( \varepsilon_1/\hbar-\omega_\xuv,\sigma_\xuv) \\
    & \times f_\fano(\epsilon_2,\bar{q}) f^*_\fano(\epsilon_1,\bar{q}) \\
\end{split}
\label{eq:reconstructed_density_matrix_Fano}
\end{equation}
where $\bar{q} = \frac{q_\text{ref} + q_\text{probe}}{2}$. This concludes the reconstruction procedure and is the one we should compare to Eq. \eqref{eq:theoretical_rho}, from which we note that the only discrepancy is the $q$ parameter which is $q_{ag}$ in the theoretical case, and $\bar{q}$ in the reconstruction. In what follows we quantify the quality of the reconstruction $\rho_H$ compared to the theoretical density matrix $\rho_\xuv$ by the fidelity $F(\rho_\xuv,\rho_H) = \text{Tr}\left(\sqrt{\sqrt{\rho_\xuv} \rho_H \sqrt{\rho_\xuv}}\right)$, or by its purity $\text{Tr}(\rho_H^2)$ compared to the expected theoretical purity $\text{Tr}(\rho_\xuv^2)$. If the last step of making the density matrix Hermitian is not taken, these two parameters acquire a small imaginary part. We discuss some of the errors in the reconstruction in Appendix D.

\section{Applications to real systems}

We illustrate the sequence on two different cases:  a photoelectron created in the vicinity of the $2s2p$ resonance in He, and one created in the unstructured continuum of Ar with two different ionic states, 3p$^5$ $^2$P$_{3/2}$ and 3p$^5$ $^2$P$_{1/2}$. \newline

\begin{figure}
    \centering
    \includegraphics[width=0.5\textwidth]{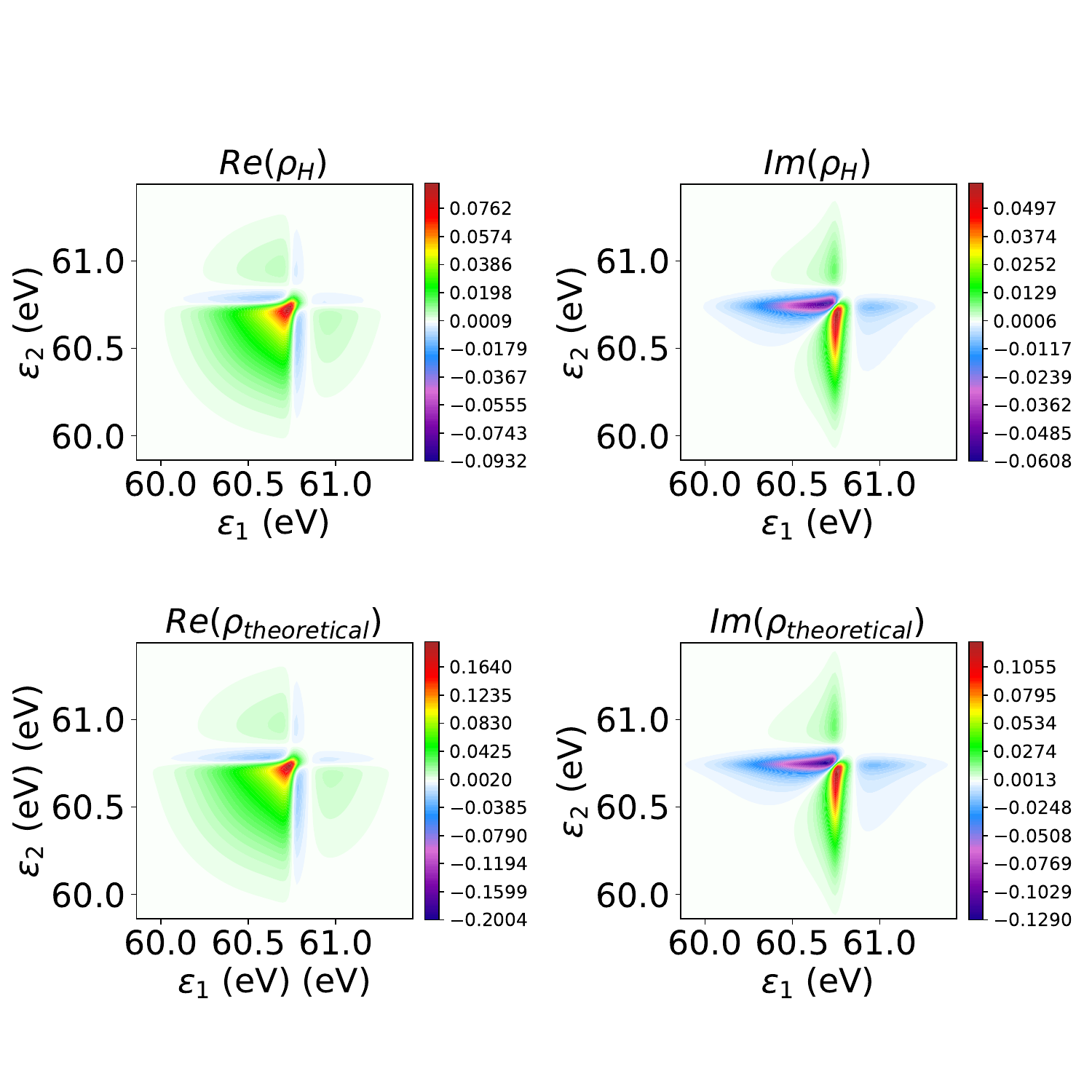}
    \caption{Reconstructed (top, eq. \eqref{eq:reconstructed_density_matrix_Fano}) and theoretical (bottom, Eq. \eqref{eq:theoretical_rho}) density matrix for He. The rainbow-KRAKEN protocol can reconstruct the density matrix with a fidelity of 0.98.}
    \label{fig:rK_Fano}
\end{figure}

\begin{figure}
    \centering
    \includegraphics[width=0.35\textwidth]{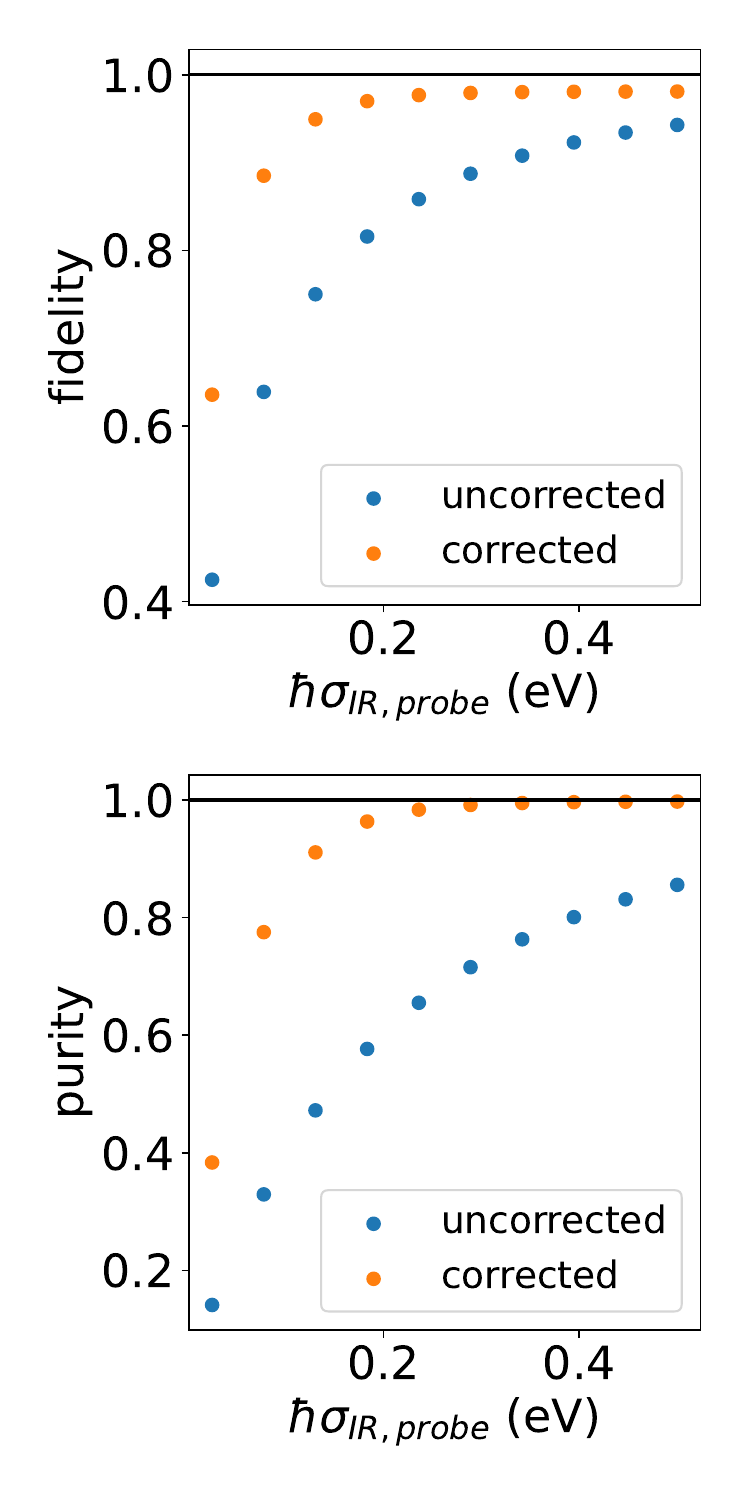}
    \caption{ (a) Fidelity and (b) purity of the He density matrix as a function of the IR probe pulse spectral width. The simulations are done for the same parameters as Figure \ref{fig:procedure}. Orange points have been corrected for finite pulse effects due to $M^{(\pm)}_{\text{probe}}$ while the blue points have not.}
    \label{fig:fidelity_and_purity_function_IR_bandwidth}
\end{figure}

\textbf{Reconstruction of a Fano resonance in He.} We simulate the reconstruction of a helium Fano resonance using $\hbar \omega_\xuv = 60.75$ eV and width $0.25$ eV. For the IR components we use $\hbar \omega_{\text{ref}} = \omega_{\text{probe}} = 1.55$ eV, and $\hbar \sigma_{\irr} = 0.001$ eV and $\hbar \sigma_{\irp} = 0.3$ eV. This corresponds to a probe pulse duration of 6 fs for a Gaussian transform-limited pulse.
We scan $\tau$ from -400 to 400 fs, and consider effective photon energies from 58 to 68 eV. 
Figure \ref{fig:rK_Fano} shows the theoretical density matrix and the reconstruction with a fidelity of $F=0.98$. We can see that the features are very well reproduced, the differences due to the functions $\Delta_\text{ref}$, $\Delta_\text{probe}$ are relatively small. 
The reconstruction is very sensitive to the IR probe bandwidth, which should cover all the energy levels we want to characterize. We can numerically explore the lower limits of an acceptable IR probe bandwidth. Figure \ref{fig:fidelity_and_purity_function_IR_bandwidth}) shows the fidelity and purity for different values of $\sigma_{\irp}$ for the case where we correct the $M^{(\pm)}_\text{probe}$ factor or for the case where we do not. We confirm that the correction is necessary to obtain reliable results, and that with the correction the values converge around $\sigma_{\irp} \approx \sigma_{\xuv}$ for the case where $\wirp = \wirr$. We also observe that while the fidelity converges to a value slightly below $F=1$, attributed to expected differences previously discussed, the purity does reach the theoretical value of 1. Thus the rainbow-KRAKEN protocol can recover the density matrix with high accuracy, and excels at measuring the purity. \newline

\textbf{Reconstruction of a mixed state.} We also investigate the reconstruction of a mixed density matrix which occurs for the states 3p$^5$ $^2$P$_{3/2}$ and 3p$^5$ $^2$P$_{1/2}$ of argon due to a spin-orbit splitting of the ionization threshold with magnitude $0.177$ meV. There is a measurement-induced decoherence due to an incomplete measurement of the degrees of freedom. Since the experiment only measures the photoelectron, the entanglement of its kinetic energy with the spin of its parent ion causes a mixed density matrix, whose purity can be affected by the XUV bandwidth \cite{Laurell2022, Laurell2023}. We can use Eqs.  \ref{eq:reconstructed_density_matrix_Fano}, \ref{eq:M_2f} in the limit $\omega_{ag} \to \infty,\; \beta = 0$ to remove the contribution of the discrete state. \newline

\begin{figure}
    \centering
    \includegraphics[width=0.5\textwidth]{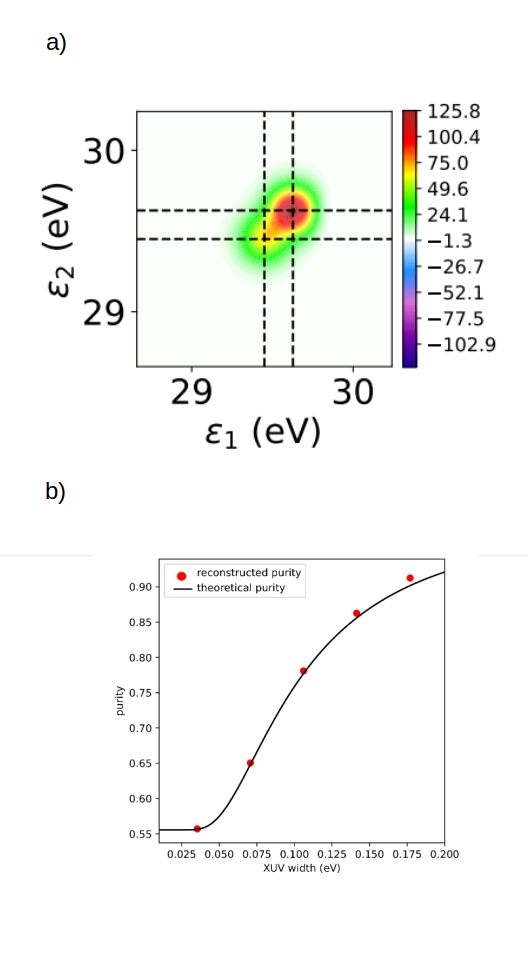}
    \caption{a) Reconstructed density matrix for Argon in the presence of spin-orbit splitting. We can clearly observe Gaussian profiles for the density matrix of each transition energy. b) Reconstructed purity as a function of XUV bandwidth (dots) compared to the theoretical purity (solid line).}
    \label{fig:reconstruction_argon}
\end{figure}

Figure \ref{fig:reconstruction_argon}.a shows the reconstruction with an XUV bandwidth small enough to resolve the spin-orbit splitting, which appear as two two-dimensional Gaussian signals. In Figure \ref{fig:reconstruction_argon}.b we show the reconstructed purity as a function of XUV bandwidth in comparison with the expected value from theory. 


\section{Conclusion}

As attosecond spectroscopy is applied to more and more complex systems, methods that can quantify the degree of decoherence and reconstruct the density matrix of photoelectrons will be needed. We have presented a pulse sequence and accompanying processing steps that can reconstruct a photoelectron density matrix in a single scan. We have shown that for systems with different structures of the excited state manifold the density matrix is reconstructed with high fidelities. The proposed experiment uses a single XUV pulse, a delayed broadband IR pulse and a temporally fixed narrowband IR pulse. The protocol makes use of shifts in Fourier space to get rid of unwanted signals as well as to label unambiguously energy levels along a second indirect dimension to the detected kinetic energy. As more and more systems begin to be studied with more discerning sequences a better picture of measurement- or vibrational-induced decoherence will emerge. 

\section{Acknowledgements}

Some of the authors acknowledge support from the Swedish Research Council (2013-8185, 2023-04603, 2023-06502), the European Research Council (advanced grant QPAP, 884900), and the Knut and Alice Wallenberg Foundation. 
HL acknowledges support from the Sweden-America Foundation.
AL is partly supported by the Wallenberg Center for Quantum Technology (WACQT) funded by the Knut and Alice Wallenberg Foundation. J.G.B. acknowledges support from Conahcyt  doctoral fellowships. D.F.S. acknowledges support from DGAPA-UNAM. 

\bibliography{Fano,RK_2}

\section*{Appendix A. Heuristic derivation of the rainbow-KRAKEN protocol} 

In order to have an intuitive understanding of the rainbow-KRAKEN protocol, we provide a heuristic derivation of the signal. The structure follows very closely to that presented in \cite{Laurell2022}. We start by considering a mixed stationary density matrix $\rho_\mathrm{xuv}$ describing the system after absorption of an XUV photon at time $\tau=0$. We describe the interaction of the system with the reference and probe IR fields as: 
\begin{align}\label{eq:rhoxuvir}
\rho_\mathrm{xuv+ir}(t)=U(t,\tau)\rho_\mathrm{xuv}U^\dagger(t,\tau),
\end{align}
where $U(t,\tau)$ is the time-ordered unitary operator:
\begin{align}\label{eq:U}
U(t,\tau)=\mathcal{T}\left[\exp\left(-i\int_{-\infty}^t dt' H(t',\tau) \right)\right],
\end{align}
where $\mathcal{T}$ is the time-ordering operator and $H(t',\tau)$ is the Hamiltonian describing the interaction with the IR pulses. The IR reference component is fixed in time at  $\tau=0$ while the IR probe component is delayed and arrives at time $\tau$. For the heuristic derivation we have set $\hbar = 1$. 

We express the Hamiltonian $H = H_0+H_{\text{light-matter}}(t)$ in the interaction picture, where the matter Hamiltonian is diagonal in the energy basis, $H_0 = \int d\varepsilon~ \varepsilon \ketbra{\varepsilon}{\varepsilon} + \int dE E \ketbra{E}{E}$. We assume a dipolar light-mater Hamiltonian $H_{\text{light-matter}} = -\mu E_{\irr}(t) -\mu E_{\irp}(t) $ where $E_{\irr}(t),\; E_{\irp}(t)$ are the electric fields of the IR pulses. We have then
\begin{equation}
\begin{split}
    H_{\text{light-matter}}'(t) &= e^{i H_0t}H_{\text{light-matter}}(t)e^{-i H_0t} \\
    &=  -\mu(t) E_{\irr}(t) -\mu(t) E_{\irp}(t).
    \label{eq:Hint}
    \end{split}
\end{equation}
where
\begin{equation}
    \mu(t) =\int \mathrm{d} E \int \mathrm{d} \varepsilon\left [ \mu_{E \varepsilon} e^{i\omega_{E \varepsilon} t} \ketbra{E}{\varepsilon} + \mu_{E \varepsilon}^* e^{-i \omega_{E \varepsilon} t} \ketbra{\varepsilon}{E} \right ] ,
\end{equation}
where $\mu_{E \varepsilon}$ are the transition dipole moment elements between states $\ket{E}$ and $\ket{\varepsilon}$. For the purposes of the derivation, we assume that the reference is a monochromatic pulse and the probe has a flat spectrum,
\begin{equation}
\begin{split}
    E_{\irr}(t) &= A_{\text{ref}}\left( e^{i \wirr t + i \phi_\text{ref}} + c.c.  \right) \\
    E_{\irp}(t) &= A_{\text{probe}}\delta(t-\tau) \left( e^{i \wirp (t-\tau) + i \phi_\text{probe}} + c.c.  \right) .
    \label{eq:EIR}
\end{split}
\end{equation}

Photoelectrons are measured at long times so that we can take the expressions in the limit $\lim_{t \to \infty} U(t, \tau)$. Then, the integration in the exponent in Eq.~\eqref{eq:U} using the expression in Eqs.~\eqref{eq:Hint}-\eqref{eq:EIR} are readily done. We write $H_{\text{light-matter}}'(t')=H_{\text{light-matter}}'^{(\text{ref})}(t') + H_{\text{light-matter}}'^{(\text{probe})}(t')$, and integrate each component separately,
\begin{equation}
\begin{split}
& \int_{-\infty}^{\infty} H_{\text{light-matter}}'^{(\text{ref})}(t') dt'  \\
     & = \int_{-\infty}^{+\infty} \mathrm{d}t' \int \mathrm{d}E \int \mathrm{d} \varepsilon 
      \left[ \mu_{E \varepsilon} e^{i\omega_{E \varepsilon} t'} \ketbra{E}{\varepsilon} \right. \\
      & \left. +\mu_{E\varepsilon}^* e^{-i\omega_{E\varepsilon} t'} \ketbra{\varepsilon}{E} \right] \\
     & \times A_{\text{ref}}\left( e^{i \wirr t' + i \phi_\text{ref}} + c.c.  \right) \\ 
    & = 2 \pi \int \mathrm{d} E \int  \mathrm{d} \varepsilon~ A_\text{ref} \mu_{E \varepsilon} \delta(\omega_{E\varepsilon} - \wirr)e^{i \phi_\text{ref}} \ketbra{E }{\varepsilon} + h.c. 
\end{split},
\end{equation}   
where we have only kept counter-rotating terms.
We can clearly see here that the role of the reference pulse is to select a given state among the $\{\ket{\varepsilon}\}$ manifold for a final detection energy $E$ obeying $E = \varepsilon + \hbar \wirr$. The integral for the probe pulse interaction Hamiltonian gives 
\begin{equation}
\begin{split}
&\int_{-\infty}^{\infty} H_{\text{light-matter}}'^{(\text{probe})}(t') dt' \\
    & = \int_{-\infty}^{+\infty} \mathrm{d}t' \int \mathrm{d}E \int \mathrm{d} \varepsilon \\
    & \times \left[ \mu_{E \varepsilon} e^{ i \omega_{E \varepsilon} t'} \ketbra{E}{\varepsilon} + \mu_{E\varepsilon}^* e^{-i \omega_{E\varepsilon} t'} \ketbra{\varepsilon}{E} \right] \\
     & \times A_{\text{probe}}\delta(t'-\tau) \left( e^{i \wirp (t'-\tau) + i \phi_\text{probe}} + c.c.  \right) \\ 
    & = \int \mathrm{d} E \int  \mathrm{d} \varepsilon~ A_{\text{probe}} \mu_{E \varepsilon} e^{i \omega_{E\varepsilon} \tau } e^{i \phi_\text{probe}} \ketbra{E }{\varepsilon} + h.c. 
\end{split}
\end{equation}   

With these expressions we can calculate the transition probability for absorbing the IR photons. We assume that $ \phi_\text{ref}=\phi_\text{probe}=0$ We first carry out a perturbative expansion of the evolution operator from Eq. \eqref{eq:U} valid for weak fields,  
\begin{equation}
\begin{split}
    \lim_{t \to \infty}  U'(t,\tau) &\approx 1 \\
   & + \left[ \int \mathrm{d}E \int \mathrm{d} \varepsilon~ \mu_{E\varepsilon} \left( 2 \pi A_{\text{ref}}\delta(\omega_{E \varepsilon} - \wirr) \right. \right. \\
   & \left. \left. \; \; \; \;+ A_{\text{probe}}e^{i\omega_{E \varepsilon} \tau} \right) + \text{h.c.} \right] + \mathcal{O}(\mu^2).
    \label{eq:approxU}
\end{split}
\end{equation}
where the prime indicates the operator in the interaction picture. 
The measured photoelectron signal for a final state with kinetic energy $E_f$ is calculated by $S_f(E_f,\tau)=\text{Tr}(\rho_\text{xuv+ir}\ketbra{E_f}{E_f})=\bra{E_f} \rho_\text{xuv+ir} \ket{E_f} $. Since the observable is a population, the expressions in the interaction and the Schr\"{o}dinger pictures are the same. 
We detect photoelectrons which result from absorption of an IR photon such that $E_f$ lies above the photoelectron energies without the absorption of an IR photon. 
Using Eq. \eqref{eq:approxU} and the fact that $\bra{E_f} \rho_\text{xuv} \ket{E_f}=0$, we have
 \begin{equation}
 \begin{split}
     & S_f(E_f,\tau) = \\ 
     & 2\pi \int \mathrm{d} \varepsilon \int \mathrm{d} \varepsilon' \mu_{E_f\varepsilon} \left( A_{\text{probe}}e^{i\omega_{E_f \varepsilon} \tau} \right)  \\ 
     & \times \rho_\text{xuv}(\varepsilon,\varepsilon') \times  
      \mu^*_{E_f\varepsilon'} \left( A_{\text{ref}}\delta(\omega_{E_f \varepsilon'} - \wirr) \right) \\
     & + 2\pi \int \mathrm{d} \varepsilon \int \mathrm{d} \varepsilon' \mu_{E_f\varepsilon} \left( A_{\text{ref}}\delta(\omega_{E_f \varepsilon} - \wirr)\right) e^{-i(\phi_\text{probe}-\phi_\text{ref} )} \\
     & \times \rho_\text{xuv}(\varepsilon,\varepsilon') \times  
     \mu^*_{E_f\varepsilon'} \left( A_{\text{probe}}e^{-i\omega_{E_f \varepsilon'} \tau} \right) \\
\end{split}
\end{equation}
 Assuming a flat, energy independent transition dipole moment $\mu_{E\varepsilon} \equiv \mu_c$ and doing the integration over $\varepsilon'$
 \begin{equation}
 \begin{split}
     &S_f(E_f,\tau) = \\ 
     & 2\pi \int \mathrm{d} \varepsilon A_{\text{probe}} A_{\text{ref}} |\mu_{c}|^2 e^{i\omega_{E_f \varepsilon} \tau} e^{i(\phi_\text{probe}-\phi_\text{ref} )} \\ 
     & \times \rho_\text{xuv}(\varepsilon,\hbar( \omega_{E_f} - \wirr)) \\
     & + 2\pi \int \mathrm{d} \varepsilon A_{\text{probe}} A_{\text{ref}} |\mu_{c}|^2 e^{-i\omega_{E_f \varepsilon'} \tau} e^{-i(\phi_\text{probe}-\phi_\text{ref} )}\\
     & \times \rho_\text{xuv}(\hbar(\omega_{E_f} - \wirr), \varepsilon) \\
\end{split}
\label{eq:heuristic_after_epsilonp}
\end{equation}
 
We Fourier transform Eq. \eqref{eq:heuristic_after_epsilonp} and interchange the integration order for $\varepsilon$ and $\tau$. We also introduce the Heaviside function $\text{Heav}(\tau)$ and extend the lower integration bound to $- \infty$. 
\begin{equation}
\begin{split}
& \tilde{S}_f(E_f,\hbar \omega_\tau) = 2\pi A_{\text{probe}} A_{\text{ref}} |\mu_{c}|^2 \\
& \times \left[  \int_{-\infty}^{+\infty} d \varepsilon \int_{-\infty}^{+\infty} d\tau  \text{Heav}(\tau) e^{-i\omega_\tau \tau + i\omega_{E_f \varepsilon} \tau} \right. \\
& \left. \times \rho_\text{xuv}(\varepsilon,\hbar( \omega_{E_f} - \wirr)) \right. \\
& \left.+ \int_{-\infty}^{+\infty} d \varepsilon \int_{-\infty}^{+\infty} d\tau  \text{Heav}(\tau) e^{-i\omega_\tau \tau - i\omega_{E_f \varepsilon} \tau} \right. \\
& \left. \times \rho_\text{xuv}(\hbar( \omega_{E_f} - \wirr),\varepsilon) \right] \\
= & \frac{\pi}{2}\int_{-\infty}^{+\infty} d \varepsilon  A_{\text{probe}} A_{\text{ref}} |\mu_{c}|^2  \\
& \times \left[ \left( \hbar \delta(\hbar\omega_\tau + \varepsilon-E_f) -\frac{i\hbar}{\pi(\hbar\omega_\tau + \varepsilon - E_f)} \right) \right. \\
&\left. \times \rho_\text{xuv}(\varepsilon,\hbar( \omega_{E_f} - \wirr)) \right. \\
& \left. + \left( \hbar\delta(\hbar\omega_\tau - \varepsilon + E_f) -\frac{i\hbar}{\pi(\hbar\omega_\tau - \varepsilon + E_f)} \right) \right. \\
& \left. \times \rho_\text{xuv}(\hbar( \omega_{E_f} - \wirr),\varepsilon) \right] \\
\end{split}
\label{eq:integral_Heaviside}
\end{equation}
When carrying out the $\varepsilon$ integral, the first term $\delta$ function is trivial to do. We consider in detail the integral
\begin{equation}
I_2 = \text{PV} \left( \int_{-\infty}^{+\infty} d\varepsilon \frac{-i \hbar}{\pi (\hbar\omega_\tau+\varepsilon-E_f)}  \rho_\text{xuv}(\varepsilon,E_f-\hbar \wirr) \right)
\label{eq:I2}
\end{equation}
where PV is the principal value of the integral. 
\begin{figure}
    \centering
    \includegraphics[width=0.5\textwidth]{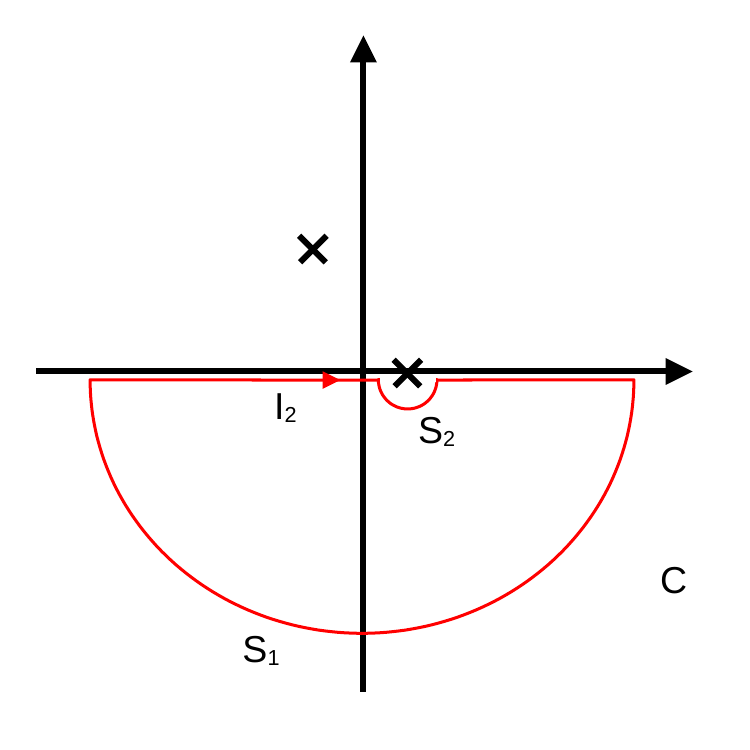}
    \caption{Contour for the evaluation of integral in Eq. \ref{eq:contours}}
    \label{fig:line_integral}
\end{figure}
We assume a well-behaved function for $\rho_\text{xuv}$, in this case with poles with respect to the integrating variable on only the upper or lower half-plane, without branch cuts and going to zero as $|\varepsilon| \to \infty$ faster than $1/\varepsilon$. To evaluate $I_2$ we close the contour $\mathcal{C}$ with a large semicircle $S_1$ and a small semicircle around the pole of the real axis $S_2$ (see Figure \ref{fig:line_integral}). 
We have:
\begin{equation}
    I_2 = \mathcal{C}-S_1-S_2
\label{eq:contours}
\end{equation}
$\mathcal{C}$ and $S_1$ vanishes and we can evaluate $S_2$ by doing a change of variables $\varepsilon+\hbar\omega_\tau-E_f = re^{i\theta}$, and evaluate the small semicircle contribution as  
\begin{equation}
    S_2 = \frac{1}{2} \lim_{r \to 0} \int_\pi^0 \frac{d\theta re^{i\theta}}{\pi re^{i\theta} }   \rho_\text{xuv}(E_f - \hbar\omega_\tau, E_f-\hbar \wirr)
\end{equation}
So,
\begin{equation}
    I_2 = \frac{1}{2} \rho_\text{xuv}(E_f - \hbar\omega_\tau,\hbar( \omega_{E_f} - \wirr))
\end{equation}
Putting together the two half-integrals and using the property $\rho(\varepsilon',\varepsilon) = \rho^*(\varepsilon,\varepsilon')$ gives the final result
\begin{equation}
\begin{split}
    \tilde{S}_f(\hbar \omega_\tau,E_f) &= A_{\text{probe}} A_{\text{ref}} |\mu_{c}|^2 \pi \\
    &\times \left[\rho_\text{xuv}( E_f - \hbar\omega_\tau ,  E_f - \hbar\wirr) \right. \\
    & \left. \times + \rho_\text{xuv}^*(E_f + \hbar\omega_\tau, E_f - \hbar\wirr) \right]
\end{split}
\end{equation}
This heuristic derivation provides an approximation to the encoding and as to the position of the signals. Redefining the new variables $\varepsilon_2^{+} =  E_f - \hbar\omega_\tau$ and $\varepsilon_1 =  E_f - \hbar\wirr$ concludes the transformation from signal to density matrix.

\section*{Appendix B. Analytical expressions of the two-photon absorption transition amplitudes}

In this appendix we derive the analytical expressions to obtain Eq. \eqref{eq:exact_analytical_expression}. The definition and meaning of the variables used here is also summarized in Table \ref{table:variables} (Appendix E).

\subsection{General expressions for a Fano structure}

For a system with an intermediate XUV manifold constituted by a discrete level at energy $\hbar \omega_{ag}$ coupled to a continuum labelled by $\{ \ket{\varepsilon} \}$, and a structurless final XUV+IR continuum (Figure \ref{fig:general_energy_levels}, left), the two-photon absorption probability amplitude for interacting with a Gaussian XUV pulse of center frequency $\omega_\xuv$ and width $\sigma_\xuv$, and a second IR pulse some delay time $\tau$ later with frequency $\omega_\ir$ and width $\sigma_\ir$ to obtain a photoelectron at energy $E$, is \cite{JimenezGalan2016}
\begin{equation}
    \mathcal{A}_{\omega_\ir}(\tau) = F(\tau)e^{i\omega_\ir \tau} \left[ w(z_E)+(\beta-\epsilon_{E_a}^{-1})(q_{ag}-i)w(z_{E_a}) \right]
    \label{eq:M_2f}
\end{equation}
Where:
\begin{equation}
\begin{split}
        F(\tau)=&-\mu_{E \varepsilon} \mu_{\varepsilon g} \pi\frac{A_\xuv A_\ir}{4\sigma_\xuv\sigma_{\text{IR}}}\\ \times& \exp \left[{-\frac{1}{2}\left(\frac{\delta^2}{\sigma^2}+\frac{\tau^2}{\sigma_t^2}+2i\frac{\sigma_{\text{IR}}}{\sigma_\xuv}\frac{\delta}{\sigma}\frac{\tau}{\sigma_t}\right)}\right]
\end{split}
\end{equation}
with $\sigma=\sqrt{\sigma_\xuv^2 +\sigma_\ir^2}$, $\sigma_t=\sqrt{\sigma_\xuv^{-2} +\sigma_\ir^{-2}}$ and $\delta=\omega_\xuv+\omega_\ir-E/\hbar$, $A_\xuv$ and $A_\ir$ are the electric field strength of the XUV and IR pulses respectively, while the complex parameter $z_E$ is defined as
\begin{equation}
   z_E=\frac{\sigma_t}{\sqrt{2}}\left[\left(\omega_\xuv-\frac{\sigma_\xuv^2}{\sigma^2}\delta-i\frac{\tau}{\sigma_t^2}\right)-E/\hbar \right] 
\end{equation}
and 
\begin{equation}
    z_{Ea}=\frac{\sigma_t}{\sqrt{2}}\left[(\omega_\xuv-\omega_{ag})+\frac{\sigma_\xuv^2}{\sigma^2}(E/\hbar-\omega_\xuv-\wirr)-i\frac{\tau}{\sigma_t^2}\right]
\end{equation}
where $q_{ag}=\frac{\mu_{\varepsilon a}}{\pi V_a \mu_{E\varepsilon}}$ is the standard Fano parameter, $\beta =  \pi \mu_{Ea}/ (V_a \mu_{E \varepsilon})$ and $\epsilon_{E_a} = (E-\hbar\omega_{ag})/\Gamma_a$. We now examine the form these equations take for the reference and probe components of the IR excitation. \newline

\subsection{Narrow bandwidth probe}

\begin{figure}[h!]
    \centering
\includegraphics[width=0.5\textwidth]{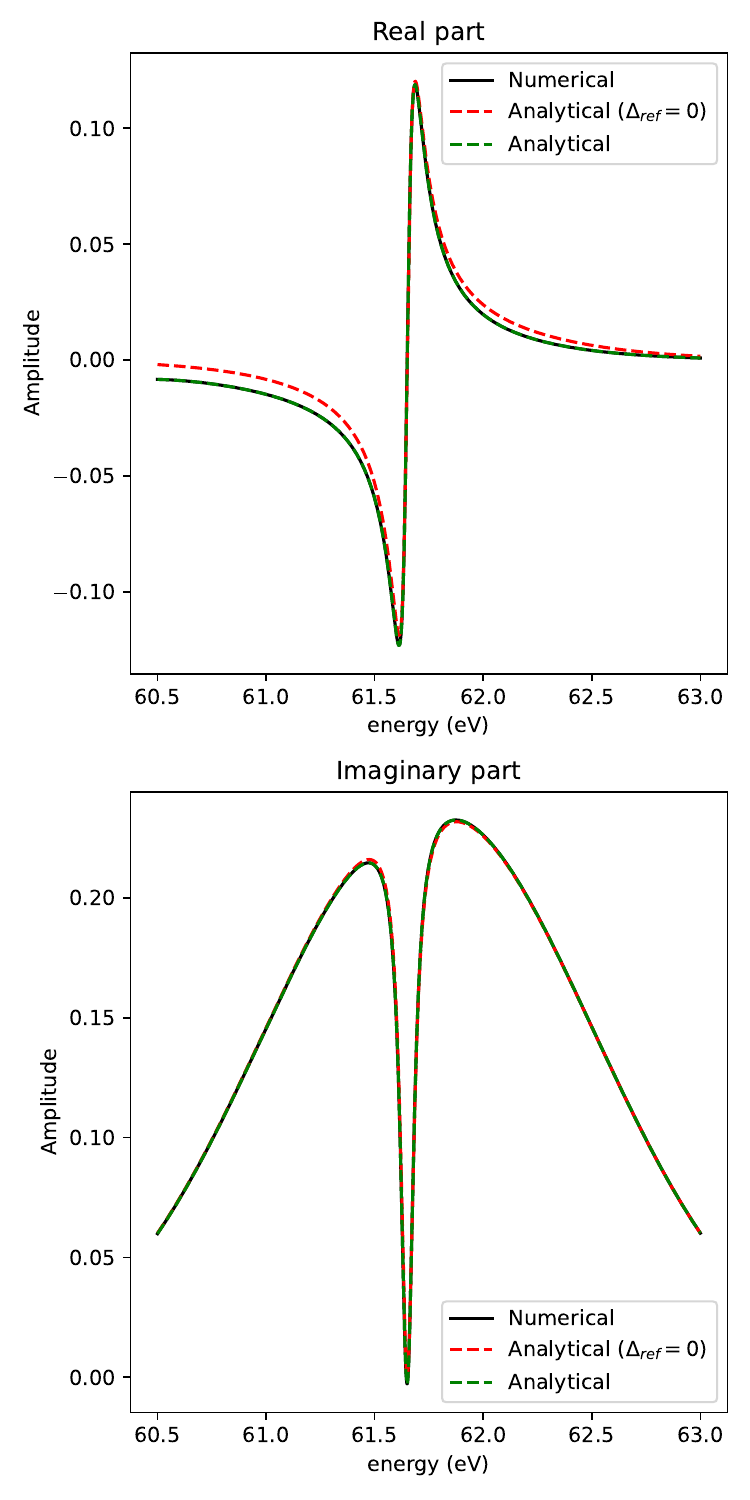}
    \caption{Simulation for a narrowband probe, its analytical limit and a weighted Fano profile. Left and right panels show real and imaginary parts, respectively. Deviations on the lower energy side become stronger due to the divergence of the $\xi$ parameter as the detected energy $E_f$ becomes comparable to $\hbar \omega_\xuv$.}
    \label{fig:narrowband_simulation}
\end{figure}

We calculate the limiting case of a standard XUV Gaussian pulse and a narrow bandwidth IR pulse. When $\lim_{\sigma_\irr \rightarrow 0} \sigma=\sigma_\xuv$ and $\lim_{\sigma_{\irr}\rightarrow 0} \sigma_t=\frac{1}{\sigma_{\irr}}$, the arguments of the Faddeeva functions have the following limiting form:
\begin{equation}
\begin{split}
 z_{E}=&-\frac{\wirr}{\sigma_\irr \sqrt{2}} \\
  z_{Ea}=&\frac{-\omega_{ag}+(E/\hbar-\wirr)}{\sigma_\irr \sqrt{2}}
 \end{split}
\end{equation}
These arguments become real and large, for which we can use an asymptotic form of the Faddeeva function, $w(z) \approx \frac{1}{-iz\sqrt{\pi}}$. We have then: 
\begin{equation}
\begin{split}
    w(z_{E})\approx& -\frac{i\sigma_\irr}{\wirr}\sqrt{\frac{2}{\pi}}
    \\
    w(z_{Ea})\approx& \frac{i\sigma_\irr}{(E/\hbar-\wirr-\omega_{ag})}\sqrt{\frac{2}{\pi}}
\end{split}
\end{equation}
and with the simplified form factor
\begin{equation}
\begin{split}
    F(0)=&-\pi \mu_{E \varepsilon} \mu_{\varepsilon g} \frac{A_\xuv A_\irr}{4\sigma_\xuv\sigma_\irr}\exp \left[{-\frac{1}{2}\left(\frac{\delta_\text{ref}^2}{\sigma_\xuv^2}\right)}\right]
\end{split}
\end{equation}
where $\delta_\text{ref} = \omega_\xuv + \omega_\irr - E/\hbar$. The two-photon transition amplitude becomes
\begin{equation}
\begin{split}
\mathcal{A}_{\wirr} (0)&=i\sqrt{2 \pi} \frac{A_\xuv A_\irr}{4\sigma_\xuv \wirr} \mu_{E \varepsilon} \mu_{\varepsilon g} \\
&\times \exp\left [-\frac{1}{2}\left(\frac{\delta_\text{ref}^2}{\sigma_\xuv^2}\right)\right]  \\
& \times \left[ 1 - \frac{\wirr (\beta-\epsilon_{E_a}^{-1})(q_{ag}-i)}{E/\hbar-\wirr-\omega_{ag}}\right]
\end{split}
\end{equation}
where $\epsilon_E = \frac{E-\hbar\omega_{ag}-\hbar\wirr}{\Gamma_a}$, and $\xi = \frac{\hbar \wirr}{\Gamma_a}\left( -\beta + \frac{1}{\epsilon_{E_a}} \right)$, $\Gamma_a = \pi V_{a}^2$. The function
\begin{equation}
    \Delta_{\text{ref}} = (q_{ag}-i)(\xi-1)
    \label{eq:Delta_ref}
\end{equation}
measures the deviations from the standard Fano profile. Its imaginary part reduces the contrast of the destructive interference point while its energy dependence slightly distorts the profile. 
We can repackage this expression closer to a Fano form
\begin{equation}
\begin{split}
\mathcal{A}_{\wirr}(0)&=i \sqrt{2 \pi}\frac{A_\xuv A_\irr}{4\sigma_\xuv \wirr} \mu_{E \varepsilon} \mu_{\varepsilon g} \\
&\times \exp \left [-\frac{1}{2}\left(\frac{\delta_\text{ref}^2}{\sigma_\xuv^2}\right)\right]  \\
& \times  \frac{(\epsilon_E + q_{ag}) + \Delta_{\text{ref}}}{\epsilon_E +i} \\
&= i \sqrt{2 \pi}\frac{A_\xuv A_\irr}{4\sigma_\xuv \wirr} \mu_{E \varepsilon} \mu_{\varepsilon g}\\
&\times G(\delta_\text{ref}, \sigma_{\xuv}) f_\text{fano}(\epsilon_E,q_{ag}+\Delta_{\text{ref}})
\end{split}
 \label{eq:analytical_sol_narrow}
\end{equation}
For the conditions $\beta \approx 0$ and the energy resonance condition $\frac{\wirr}{E-\omega_{ag}} \approx 1$, we have that $\xi \approx 1$, $\Delta_{\text{ref}} \approx 0$ and the transition probability amplitude is proportional to the Fano profile weighted by the XUV bandwidth (see Figure \ref{fig:narrowband_simulation}). \newline
\begin{equation}
\mathcal{A}_{\wirr}(0) \propto \frac{(\epsilon_E + q_{ag})}{\epsilon_E +i} \exp\left [-\frac{1}{2}\left(\frac{\delta_\text{ref}^2}{\sigma_\xuv^2}\right)\right]   
\end{equation}

\subsection{Fourier transform of the two-photo absorption for a broad bandwidth probe}

The use of a broadband probe does not lead to a simplifying limit as in the narrowband reference case, and the Fourier transform of the entire expression needs to be calculated. We rewrite
\begin{equation}
\begin{split}
    z_E &= -i\frac{\tau+i\tau_E}{\sqrt{2}\sigma_t} \\
    z_{E_a} &= -i\frac{\tau+i\tau_a}{\sqrt{2}\sigma_t} \\
\end{split}
\end{equation}
where $\sigma = \sqrt{ \sigma_\xuv^2 + \sigma_\irp^2 }$, $\sigma_t = \sqrt{ \sigma_\xuv^{-2} + \sigma_\irp^{-2} }$, 
$\tau_E = \sigma_t^2 \delta_E$ and $\tau_a =  \sigma_t^2 \delta_a$, $\delta_E=\omega_\xuv-E/\hbar-\frac{\sigma_\xuv^2}{\sigma^2}\delta$ and $\delta_a=\omega_\xuv-\omega_{ag}-\frac{\sigma_\xuv^2}{\sigma^2}\delta$. Their squares are
\begin{equation}
    \begin{split}
        (z_{E})^2 &= -\frac{\tau^2}{2 \sigma_t^2} -i\frac{\tau \tau_E}{ \sigma_t^2} + \frac{\tau_E^2}{2\sigma_t^2} \\
        (z_{E_a})^2 &= -\frac{\tau^2}{2 \sigma_t^2} -i\frac{\tau \tau_a}{\sigma_t^2} + \frac{\tau_a^2}{2\sigma_t^2} \\
    \end{split}
\end{equation}

\begin{figure}
    \centering
    \includegraphics[width=0.5\textwidth]{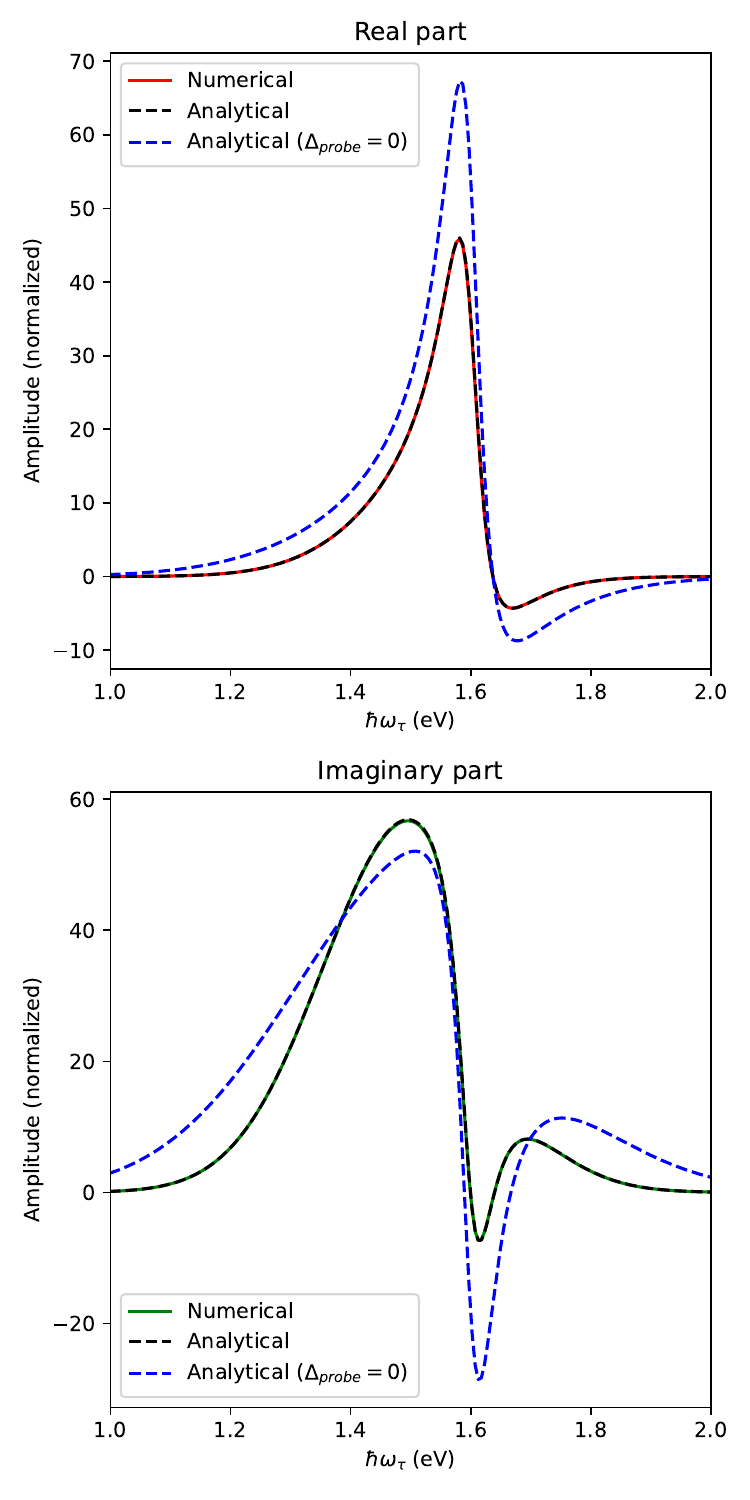}
    \caption{Numerical simulation and analytical result of the Fourier transform $\tilde{\mathcal{A}}_{\text{ir,probe}}(\omega_\tau)$.}
    \label{fig:analytical_sim_FT_verification}
\end{figure}

We decompose the form factor as:
\begin{equation}
    \begin{split}
        \mathcal{F} (\tau) & = f_0 f_{\phi}(\tau) f_E(E_f) f_G(\tau) 
    \end{split}
\end{equation}
where $f_0 = -\pi \mu_{E\varepsilon} \mu_{\varepsilon g} \frac{A_\xuv A_\irp}{4 \sigma_\xuv \sigma_\irp}$, $f_{\phi}(\tau) = e^{-i \omega_F \tau}$, $f_E(E_f) = e^{-\frac{\delta^2}{2\sigma^2}}$, $f_G(\tau) = e^{-\frac{\tau^2}{2\sigma_t^2}}$ and 
 $\omega_F = \frac{\sigma_{\irp}}{\sigma_{\xuv}}\frac{\delta}{\sigma \sigma_t}$. Writing $w(z) = e^{-z^2}(1-\erf(-iz))$ we can simplify Equation \eqref{eq:M_2f} as
\begin{equation}
\begin{split}
\mathcal{A}_{\wirp}(\tau) &= f_0 f_E(E_f) \\
& \times \sum_{j=E,a} A_{j} e^{-i \Omega_j \tau} e^{-\frac{\sigma_t^2 \delta_j^2}{2}}(1-\erf(-iz_j))
\label{eq:compact_form_M2f}
\end{split}
\end{equation}
where $A_E=1$, $A_a=(\beta-\epsilon_{E_a}^{-1})(q_{ag}-i)$, $\Omega_j = \omega_F-\delta_j-\wirp$ and $z_j = \frac{-i\tau + \tau_j}{\sqrt{2}\sigma_t}$, where $j=E,a$ corresponds to the direct transition (E) or that going through the discrete state (a). \newline
\
We can make the simplifying assumption that for the range of parameters relevant for the experiment $(1-\erf(-iz_j) ) \approx -\erf(-iz_j)$. 
The Fourier transform of the error function is
\begin{equation}
    F_{\tau} \left\{\erf(\tau) \right \} = -i \frac{e^{-\omega_\tau^2/4}}{\omega_\tau/2}
\end{equation}
and so using the Fourier identities for rescaling and shifts $F_\tau(e^{-i \Omega \tau}f(\tau/a+b)) = \abs{a}\tilde{f}(a(\Omega+\omega_\tau))e^{iab(\Omega+\omega_\tau)}$ where $F_\tau(f(\tau))=\tilde{f}(\omega_\tau)$:
\begin{equation}
\begin{split}
    & F_{\tau}\left \{e^{-i\Omega_j \tau}\erf \left(-\frac{\tau}{\sqrt{2}\sigma_t}-i\frac{\sigma_t}{\sqrt{2}}\delta_j \right) \right \} \\
    & = 2i \frac{e^{-\frac{(\omega_\tau + \Omega_j + \delta_j)^2\sigma_t^2}{2}}e^{\frac{\sigma_t^2 \delta_j^2}{2}}}{\omega_{\tau}+\Omega_j}
    \end{split}
\end{equation}
The Fourier transform of eq. \eqref{eq:compact_form_M2f} is:
\begin{equation}
    \begin{split}
        \tilde{\mathcal{A}}_{\wirp}(\omega_\tau) & = - 2i f_0 \\
        &\times e^{-\frac{\delta^2}{2\sigma^2}} e^{-\frac{\sigma_t^2}{2}(\omega_{\tau}-\wirp+\frac{\sigma_\irp^2}{\sigma^2}\delta)^2} \\
        & \times \left[ \frac{1}{\omega_\tau} + \frac{ (\beta-\epsilon_{E_a}^{-1})(q_{ag}-i)}{\omega_\tau-(E-\tilde{\omega}_{ag})} \right] \\
    \end{split}
    \label{eq:Aprobe}
\end{equation}
The expression in Eq. \eqref{eq:Aprobe} can be decomposed into a set of prefactors (first line), a modulation function that determines the region where the signal will appear, depending on the properties of the pulse sequence (second line), and an energy dependence reflecting the structure of the energy levels (third line). 
Defining
\begin{equation}
    M^{(+)}_{\text{probe}}(\omega_\tau) = \frac{1}{\pm\omega_\tau} e^{-\frac{\delta^2}{2\sigma^2}} e^{-\frac{\sigma_t^2}{2}(\omega_{\tau}-\wirp+\frac{\sigma_\irp^2}{\sigma^2}\delta)^2}
\end{equation}
Remembering that $\epsilon_{E_a} = (E-\omega_{ag})/\Gamma_a$ and $\tilde{\omega_{ag}} = \omega_{ag} -i \Gamma_a$ with $\Gamma_a=\pi V_{a}^2$, we have:
\begin{equation}
    \begin{split}
        \tilde{\mathcal{A}}_{\wirp}(\omega_\tau) & = -2 i f_0 
        M^{(+)}_{\text{probe}}(\omega_\tau) \\
        & \times \left[ \frac{(-\epsilon_{\tau}+q_{ag}) + \Delta_{\text{probe}} }{(-\varepsilon_{\tau}+i)} \right] \\
        & = - 2i f_0 
        M^{(+)}_{\text{probe}}(\omega_\tau) \\
        & \times f_\text{fano}(-\epsilon_\tau, q_{ag} + \Delta_{\text{probe}}) 
        \label{eq:analytical_sol_FT}
    \end{split}
\end{equation}
where the function that measures the deviation from the Fano profile 
\begin{equation}
\Delta_{\text{probe}} = (q_{ag}-i)\left(\xi \frac{\omega_{\tau}}{\wirr}-1 \right)
\label{eq:Delta_probe}
\end{equation}
and $\xi = \frac{\hbar \wirr}{\Gamma_a}\left(\frac{1}{\epsilon_{E_a}}-\beta \right)$, and we have defined a new detuning parameter $\epsilon_\tau \equiv \epsilon_{\tau}^{(+)} = \frac{\hbar \omega_\tau - (E_f - \hbar \omega_{ag})}{\Gamma_a}$. If we can make the approximations $\Delta_{\text{probe}} \approx 0$ then the profile has the Fano form. The comparison for a numerical Fourier transform, the analytical solution of Eq. \eqref{eq:analytical_sol_FT} and the simplification for $\Delta_{\text{probe}} = 0$ is shown in Figure \ref{fig:analytical_sim_FT_verification}. 
We need to multiply the analytical expression by a factor of $\pi^2/2$ to agree with the numerical Fast Fourier Transform routine in Python.

\subsection{Building the density matrix}

From the expressions for narrowband and broadband pulses we can construct a density matrix. Using equation \eqref{eq:total_signal_time} we have:

\begin{equation}
\begin{split}
\tilde{S}_{RK}(E_f,\omega_\tau) &=  -i \sqrt{2 \pi}\frac{A_\xuv A_\irr}{4\sigma_\xuv \wirr} \mu_{E \varepsilon} \mu_{\varepsilon g}\\
&\times G(\delta_\text{ref}, \sigma_{\xuv}) f^*_\text{fano}(\epsilon_E,q_{ag}+\Delta_{\text{ref}}) \\
&\times (- 2i f_0 
M^{(+)}_{\text{probe}}(\omega_\tau) \\
& \times f_\text{fano}(-\epsilon_\tau, q_{ag} + \Delta_{\text{probe}})) \\
& + i \sqrt{2 \pi}\frac{A_\xuv A_\irr}{4\sigma_\xuv \wirr} \mu_{E \varepsilon} \mu_{\varepsilon g}\\
&\times G(\delta_\text{ref}, \sigma_{\xuv}) f_\text{fano}(\epsilon_E,q_{ag}+\Delta_{\text{ref}}) \\
&\times    2i f_0 
M^{(-)}_{\text{probe}}(\omega_\tau) \\
& \times f_\text{fano}^*(-\epsilon_\tau, q_{ag} + \Delta_{\text{probe}})
\end{split}     
\end{equation}
which we can write compactly as:
\begin{equation}
\begin{split}
    \tilde{S}_{RK}(E_f,\omega_\tau) &= \tilde{S}_{RK}^{(+)}(E_f,\omega_\tau) + \tilde{S}_{RK}^{(-)}(E_f,\omega_\tau), \\
    \tilde{S}_{RK}^{(+)}(E_f,\omega_\tau) & =
    I_0 G(\delta_{\text{ref}}, \sigma_\xuv) M_{\text{probe}}^{(+)}(E_f,\omega_\tau) \\
    & \times f_\fano(\varepsilon_{E_f}, q_{\text{ref}}) f^*_\fano(-\varepsilon_\tau, q_{\text{probe}}), \\
    \tilde{S}_{RK}^{(-)}(E_f,\omega_\tau) & = I_0 G(\delta_{\text{ref}}, \sigma_\xuv) M_{\text{probe}}^{(-)}(E_f,\omega_\tau) \\
    & \times f^*_\fano(\varepsilon_{E_f}, q_{\text{ref}}) f_\fano(\varepsilon_\tau, q_{\text{probe}})
\end{split}
\label{eq:exact_analytical_expression_SI}
\end{equation}
where $I_0 = -2\sqrt{2 \pi}\frac{A_\xuv A_\irr}{4\sigma_\xuv \wirr} \mu_{E \varepsilon} \mu_{\varepsilon g} f_0$. 

\section*{Appendix C. Arbitrary energy structure}

\textbf{Multiple discrete levels in the XUV manifold}. The arguments outlined earlier apply for an intermediate manifold (after absorption of an XUV photon) that consists of a discrete level (resonance) coupled to a continuous manifold of levels, and a final manifold (after subsequent absorption of an IR photon) that consists of only a continuous manifold of levels. We would like to treat the arbitrary case of an arbitrary number of discrete levels in the intermediate manifold. The two-photon cross-section is then $\mathcal{A}_{\omega_2}(\tau) = F(\tau)e^{i\omega_2 \tau} \left[ w(z_E)+\sum_j (\beta_j -\varepsilon_{E_{a_j}}^{-1})(q_{a_jg}-i)w(z_{E_{a_j}}) \right]$. We need to calculate the two terms that make up the final expression. These can be directly written down as:

\begin{equation}
\begin{split}
\mathcal{A}_{\wirr} (0)&=i\sqrt{2 \pi} \frac{A_\xuv A_\irr}{4\sigma_\xuv \wirr} \mu_{E \varepsilon} \mu_{\varepsilon g} \\
&\times \exp\left [-\frac{1}{2}\left(\frac{\delta_\text{ref}^2}{\sigma_\xuv^2}\right)\right]  \\
& \times \sum_i \left[ 1 - \frac{\wirr (\beta-\epsilon_{E_{a_i}}^{-1})(q_{a_ig}-i)}{E/\hbar-\wirr-\omega_{a_ig}}\right]
\end{split}
\end{equation}

and 

\begin{equation}
    \begin{split}
        \tilde{\mathcal{A}}_{\wirp}(\omega_\tau) & = -i 2 f_0 e^{-\frac{\delta^2}{2\sigma^2}} e^{-\frac{\sigma_t^2}{2}(\omega_{\tau}-\omega_2+\frac{\sigma_2^2}{\sigma^2}\delta)^2} \\
        & \times \left[ \frac{1}{\omega_\tau} + \sum_j \frac{ (\beta_j-\epsilon_{E_{a_j}}^{-1})(q_{a_jg}-i)}{\omega_\tau-(E_f-\tilde{\omega}_{a_jg})} \right] \\
    \end{split}
\end{equation}

Having multiple levels coupled to the continuum introduce deviations from the Fano profile and destroy a perfect destructive interference, but the density matrix can nonetheless be reconstructed either exactly by assuming a model of the energy structure or model-free with a small error. \newline

\textbf{Multiple discrete levels in the XUV manifold and multiple discrete levels in the XUV+IR manifold}. In a more general setting we can have discrete levels $\ket{a}$ in the XUV manifold, as well as discrete levels $\ket{b}$ in the XUV+IR manifold, as well as discrete levels $\ket{n}$ accessible by photon absorption from the ground state that are not coupled to ionized states. 
We use the expression for the two-photon transition amplitude from Jimenez-Galan et al. \cite{JimenezGalan2016}, 
\begin{eqnarray}
\begin{split}
    &A_{\omega_{ir,probe}}(\tau)=F(\tau)e^{i\omega_{ir,probe}\tau}
   \frac{\epsilon_{Eb}+i}{\epsilon_{Eb}-i}
   \\
   &\times \left[\frac{\epsilon_{Ea}+q_{\tilde{a}g}}{\epsilon_{Ea}+i}
    w(z_E)+(q_{\tilde{a}g}-i)w(z_{\tilde{Ea}})
    \right.
    \\
    &\times \left. \left(\beta\frac{\epsilon_{Eb}+q_{\tilde{b}a}}{\epsilon_{Eb}+i}-\frac{1}{\epsilon_{Ea}+i}+\frac{\delta_{ba}(q_{\tilde{a}b}-i)-\xi_{ba}}{\epsilon_{Eb}+i}\right)
    \right.
    \\
    &+\left.
    \sqrt{\frac{2}{\pi}}\frac{1}{\sigma_t}\frac{\mu_{b\varepsilon}}{\mu_{E \varepsilon}(\epsilon_{Eb}+i)} \right. \\
    &\left. + \sum_{n}\frac{\epsilon_{Eb}+q_{\tilde{b}n}}{\epsilon_{Eb}+i}\frac{\mu_{En}\mu_{n g}}{\mu_{E \varepsilon} \mu_{\varepsilon g}}w(z_{En})\right] \label{eq:A(t)General}
    \end{split}
\end{eqnarray}
where we have introduced new parameters
\begin{eqnarray}
\begin{split}
\delta_{ba}=\frac{\Gamma_a/2}{V_{bE}}\frac{\mu_{b\varepsilon}}{\mu_{E \varepsilon} }; \; \; \xi_{ba}=\frac{V_{a\varepsilon}}{V_{bE}}\frac{\mu_{ba}}{\mu_{E\varepsilon}}
\end{split}
    \end{eqnarray}
To construct the expression of a narrow bandwidth IR pulse and XUV pulse, we use the same limiting forms of the error function and obtain
\begin{eqnarray}
    \begin{split}
&\mathcal{A}_{\wirr} (0)=i\sqrt{2 \pi} \frac{A_\xuv A_\irr}{4\sigma_\xuv \wirr} \mu_{E \varepsilon} \mu_{\varepsilon g} \\
&\times \exp\left [-\frac{1}{2}\left(\frac{\delta_\text{ref}^2}{\sigma_\xuv^2}\right)\right]\frac{\epsilon_{Eb}+i}{\epsilon_{Eb}-i} 
\\
&\times
\left[\frac{\epsilon_{Ea}+q_{\tilde{a}g}}{\epsilon_{Ea}+i}-
\frac{(q_{\tilde{a}g}-i)\hbar\wirr}{E-\hbar\wirr-\hbar\omega_{\tilde{a}g}} 
\right.
\\
& \left. \times
\left(\beta\frac{\epsilon_{Eb}+q_{\tilde{b}a}}{\epsilon_{Eb}+i}-\frac{1}{\epsilon_{Ea}+i}+\frac{\delta_{ba}(q_{\tilde{a}b}-i)-\xi_{ba}}{\epsilon_{Eb}+i}\right) \right.
\\
&\left. +\frac{\wirr}{i}\frac{\mu_{b\varepsilon}}{\mu_{E \varepsilon}(\epsilon_{Eb}+i)} \right.
\\
&\left.-\sum_{n}\frac{\epsilon_{Eb}+q_{\tilde{b}n}}{\epsilon_{Eb}+i}\frac{\mu_{En}\mu_{n g}}{\mu_{E \varepsilon} \mu_{\varepsilon g}}\frac{\hbar\wirr}{E-\hbar\wirr-\hbar\omega_{ng}} \right]
    \end{split}
\end{eqnarray}
and for the Fourier transform of the broadband probe
\begin{equation}
    \begin{split}
  \tilde{\mathcal{A}}_{\wirp}(\omega_\tau) & = -2i f_0  M^{(+)}_{\text{probe}}(\omega_\tau) \frac{\epsilon_{Eb}+i}{\epsilon_{Eb}-i} \\
        & \times \left[
        \frac{\epsilon_{Ea}+q_{\tilde{a}g}}{\epsilon_{Ea}+i}+\frac{\hbar\omega_\tau(q_{\tilde{a}g}-i)}{\hbar\omega_\tau-(E-\tilde{\omega}_{ag})}\right.
\\
& \left. \times \left(\beta\frac{\epsilon_{Eb}+q_{\tilde{b}a}}{\epsilon_{Eb}+i}-\frac{1}{\epsilon_{Ea}+i}+\frac{\delta_{ba}(q_{\tilde{a}b}-i)-\xi_{ba}}{\epsilon_{Eb}+i}\right) \right.
\\
& \left. + \frac{i\omega_\tau\mu_{b\varepsilon}}{\mu_{E \varepsilon}(\epsilon_{Eb}+i)}\right.
\\
& \left. +\sum_n\frac{\epsilon_{Eb}+q_{\tilde{b}n}}{\epsilon_{Eb}+i}\frac{\mu_{En}\mu_{n g}}{\mu_{E \varepsilon} \mu_{\varepsilon g}}
\frac{\hbar\omega_\tau}{\hbar\omega_\tau-(E-\hbar\omega_{ng})}\right]
    \end{split}
\end{equation}
    
\section*{Appendix D. Sources of error}

We now discuss the sources of error that can lead to an imperfectly reconstructed density matrix. \newline



\textit{Model free vs. fitted reconstruction of the density matrix.} Ideally, the rainbow-KRAKEN protocol reconstructs the density matrix model-free, that is, faithfully provide $\rho_{\xuv}$ after the transformations to the signal of the previous section without assumptions on the energy level or structure. As becomes apparent from Eq. \eqref{eq:exact_analytical_expression}, we incur
in an error described by the functions $\Delta_{\text{probe}}$ and $\Delta_{\text{ref}}$. Model-free reconstructions constitute the approximation that $\Delta_{\text{probe}}=\Delta_{\text{ref}}=0$ (more complex energy structures will also have equivalent error functions as are calculate in Appendix C.). This approximation, however, is not drastic and we obtain fidelities close to 0.98, meaning that deviations introduce a negligible error. This error is absent when determining the purity. \newline

\begin{figure}
    \centering
    \includegraphics[width=0.5\textwidth]{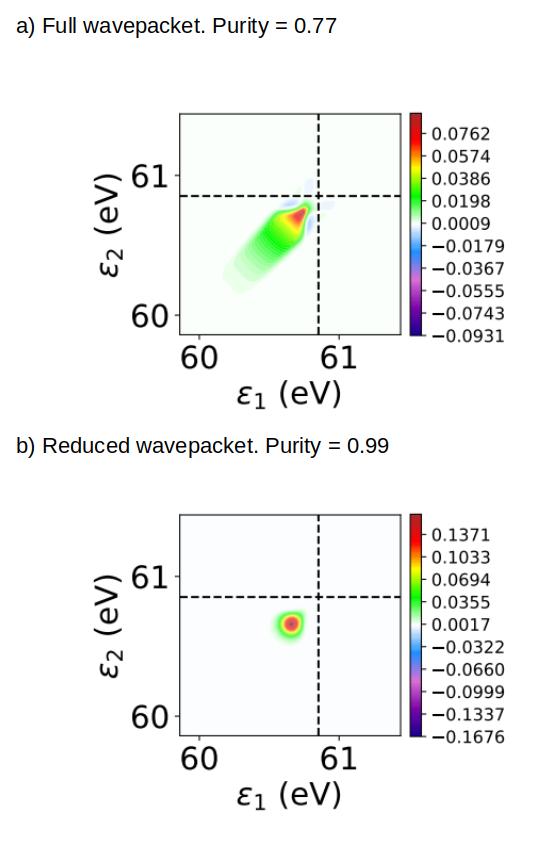}
    \caption{a) Reconstruction of the density matrix for He using $\hbar \sigma_\irp = 0.05$ eV. b) Reconstruction using a mask corresponding to $\sigma_\text{eff} = \sigma_\irp$. }
    \label{fig:reduce_sigma_1}
\end{figure}

\textit{Finite IR probe pulse bandwidth}. For a narrower IR bandwidth than can be corrected for by \textit{Step 3}, we can only reconstruct a portion of the density matrix close to the diagonal (Fig. \ref{fig:reduce_sigma_1}.a). In itself, this would introduce an erroneous degree of decoherence between levels farther apart than the IR bandwidth. 
There are two options to solve this: we can shift the probe spectrum to sit on the edge of the IR reference frequency and so extend the energy distance between states whose coherence can be probed. All of the expressions derived apply, except that then only an upper or lower triangular part of the density matrix can be reconstructed. However since it is Hermitian the remaining part can be carefully inferred. Very large density matrices can be reconstructed by parts by shifting the IR probe spectrum with respect to the IR  reference frequency.
The other option is to artificially reduce the XUV bandwidth to focus our attention on only a part of the wavepacket. This can be achieved by multiplying the x-axis by $(G( \delta_\text{ref} ,\sigma_\xuv)+\zeta)^{-1}(G( \delta_\text{ref} ,\sigma_{\text{eff}}))$, where $\sigma_{\text{eff}}<\sigma_\xuv$ is an effective wavepacket width that can be reconstructed by the IR probe pulse, and modifying the correction to the $M^{(\pm)}$ function to similarly restrict the XUV wavepacket. 
It is an identical normalization in structure as in \textit{Step 3} except that instead of increasing the bandwidth - which can only be done so far - we reduce it. 
An analysis of the fidelity and purity as a function of IR probe bandwidth in the case where $\wirr = \wirp$ shows that a convergence is reached when $\sigma_\irp \approx \sigma_\xuv$ (Figure \ref{fig:fidelity_and_purity_function_IR_bandwidth}). Figure \ref{fig:reduce_sigma_1}.b shows the restricted bandwidth reconstruction where a purity of 0.99 is obtained.  \newline

\begin{figure}
    \centering
    \includegraphics[width=0.5\textwidth]{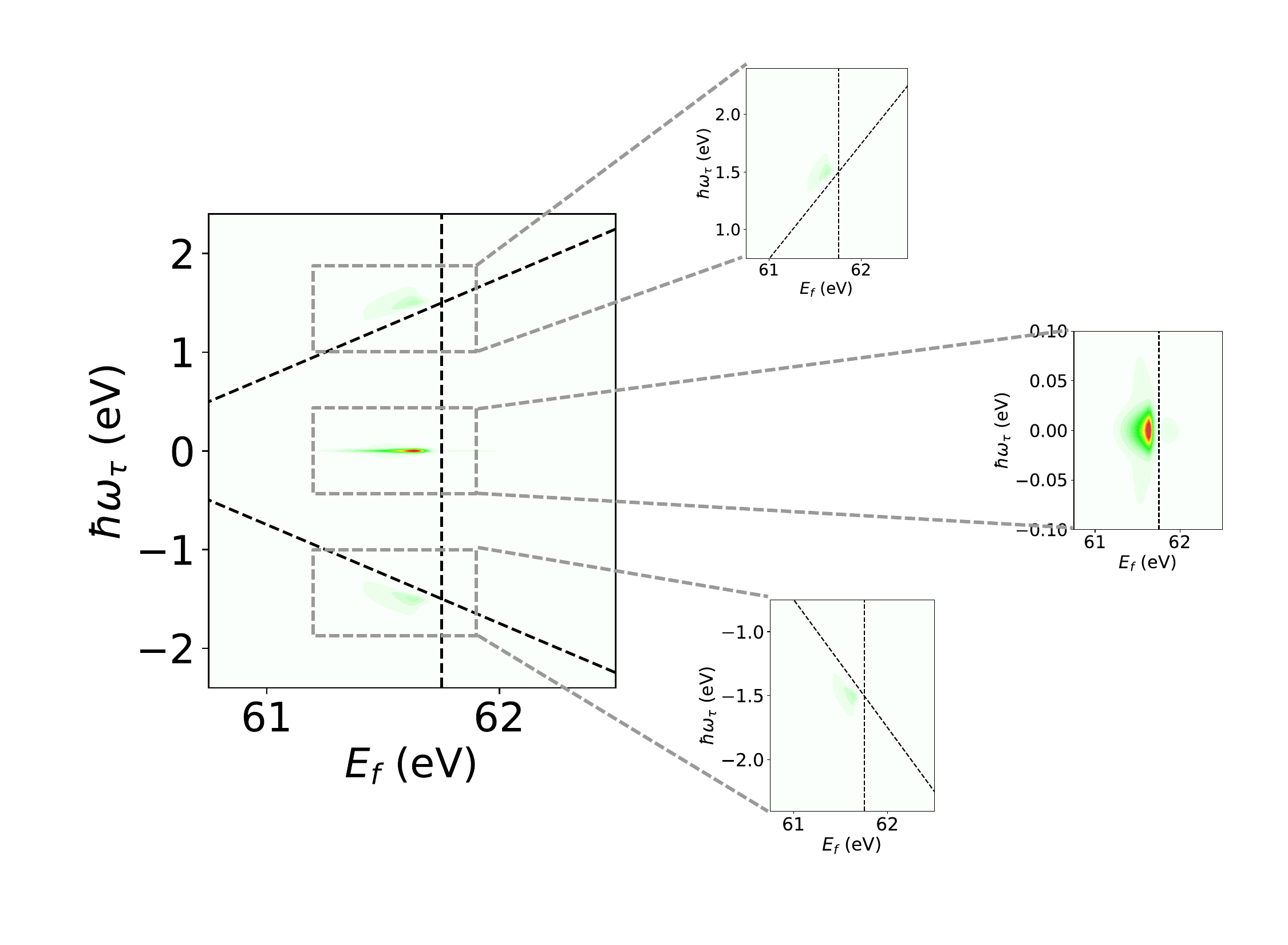}
    \caption{Fourier encodings of all of the photoelectron signals coming from interactions with the IR probe, the IR reference, and their interference.}
    \label{fig:all_Fourier_encodings}
\end{figure}

\textit{Impartial substractions of pure probe and pure reference contributions}. The signal calculated in Eq.\eqref{eq:total_signal_time} relies on substracting the signal where the photoelectron wavepacket interacts only with the reference and only with the probe. Typically, substractions are imperfect, leading to errors in the final signal. However, these parasitic contributions appear clustered within $\sigma_\irp$ of $\omega_\tau=0$, distinctly separated from the regions where the desired signal appears. Figure \ref{fig:all_Fourier_encodings} shows the Fourier transform of the interferogram without substraction of the parasitic contributions. We can see the positive and negative frequency signal that encode the density matrix, and at zero frequency very far from the desired signal all of the contributions that we do not want. Lock-in modulation of the IR components can automatically remove the photoelectrons arising solely from the reference or probe components, however the sequence naturally isolates the signals in separate places of Fourier space, making it truly a single-scan density matrix reconstruction. As a technical note, since the reference IR probe is not scanned, there will be a constant signal for all delay times. Fourier transforming a constant signal with a finite time-delay will generate artificial high-frequency terms, so that if Fourier filtering is used to remove the unwanted contributions a windowing function is needed to make the signal go to zero towards the end of the scanning range. 

\section*{Appendix E. Nomenclature}

\begin{widetext}
\centering
\begin{table*}[h!]
\centering
\begin{tabular}{|c|l|c|}
\hline
Variable & Meaning & Units \\
\hline
$\omega_{\xuv}$ & Freq. of the XUV photon & rad/s \\
$\sigma_{\xuv}$ & Width of the XUV photon & rad/s \\
$\omega_{\irr}$ & Freq. of the IR reference photon & rad/s \\
$\sigma_{\irr}$ & Width of the IR reference photon & rad/s \\
$\omega_{\irp}$ & Freq. of the IR probe photon & rad/s \\
$\sigma_{\irp}$ & Width of the IR probe photon & rad/s \\ \hline
$\ket{g}$ & ground state & \\
$\{ \ket{\varepsilon} \}$ & continuous manifold of states accessible after absorption of an XUV photon & \\
$\ket{\varepsilon_1},\; \ket{\varepsilon_2},\; \ket{\varepsilon_{2'}} $ & levels within $\{ \ket{\varepsilon} \}$ & \\
$\{ \ket{E} \}$ & continuous manifold of states accessible after absorption of an XUV and an IR photon & \\
$\ket{E_f}$ & level within $\{ \ket{E} \}$ & \\
$\ket{a}$ & discrete level accesssible by absorption of an XUV photon & \\ \hline
$ \tilde{\omega}_{ag} $ & $\omega_{ag}-i\Gamma_a/\hbar$ & rad/s \\
$\hbar \omega_{ag}$ & transition energy from the ground state to $\ket{a}$ & eV \\
$V_a$ & electronic coupling between $\ket{a}$ and $\ket{\varepsilon}$ & eV \\
$\Gamma_a$ & $\pi \abs{V_a}^2$ & eV \\ \hline
$\tau$ & Delay between the XUV and IR probe pulses & fs \\
$\omega_{\tau}$ & Conjugate freq. to the delay $\tau$ & rad/s \\ \hline
$\delta_i$ & $\omega_\xuv + \omega_\text{IR,i} - E_f/\hbar$, for $i=$ ref, probe & rad/s \\ 
$\delta_E$ & $\omega_\xuv-E/\hbar-\frac{\sigma_\xuv^2}{\sigma^2}\delta$ & rad/s \\ 
$\delta_a$ & $\omega_\xuv-\omega_{ag}-\frac{\sigma_\xuv^2}{\sigma^2}\delta$ & rad/s \\ 
$\delta_{\omega_{\tau}}^{(\pm)}$ &  $\pm \omega_{\tau}+ \omega_\xuv-E_f/\hbar$ & rad/s \\ 
$\xi$ & $\frac{\hbar \wirr}{\Gamma}\left(\frac{1}{\epsilon_{E_a}}-\beta \right)$ & \\
$\Delta_{\text{ref}}$ & $(q-i)(\xi -1)$ & \\
$\Delta_{\text{probe}}$ & $(q-i)(\xi \frac{\omega_{\tau}}{\wirr}-1)$ & \\

\hline
\end{tabular}
\caption{Table of variables and their meanings with corresponding units.}
\label{table:variables}
\end{table*}
\end{widetext}

\end{document}

\section{Appendix C. Experimental realization}

The final signal measured this way will also have contributions from the photoelectron interacting with the reference alone, but this will appear as a DC signal at $\omega_\tau=0$ and can be removed by focusing on the signal around $\pm \wirp$. The signal can also be measured without a chopper as the signal of the photoelectron interacting only with the probe will also appear around $\omega_{\tau}$ with a width of $\sigma_{\text{IR,probe}}$, which is much smaller than $\wirp$.


\end{document}